\documentclass[useAMS,usenatbib,usegraphicx,uselongtable]{mn2e}
\usepackage[varg]{txfonts}

\usepackage{longtable}
\usepackage{lscape}
\usepackage{graphicx}
\usepackage{rotating}
\usepackage{hyperref}

\title[Very low-luminosity Class I/Flat outflow sources]{Very low-luminosity Class I/Flat outflow sources in $\sigma$ Orionis}

\author[Riaz et~al.]
{B. Riaz,$^{1}$\thanks{E-mail: basriaz@astro.umd.edu} M. Thompson,$^{1}$ E. T. Whelan,$^{2}$ N. Lodieu$^{3,4}$     \\
$^{1}$Centre for Astrophysics Research, Science \& Technology Research Institute, University of Hertfordshire, Hatfield, AL10 9AB, UK \\
$^{2}$Institute f\"{u}r Astronomie und Astrophysik, Eberhard Karls University Tuebingen, Sand 1, D-72076 T\"{u}bingen, Germany \\
$^{3}$Instituto de Astrofisica de Canarias, E-38206 La Laguna, Tenerife, Spain \\
$^{4}$Departamento de Astrof\'isica, Universidad de La Laguna, E-38206 La Laguna, Tenerife, Spain }

\begin{document}

\date{}

\pagerange{\pageref{firstpage}--\pageref{lastpage}} \pubyear{2014}

\maketitle

\label{firstpage}

\begin{abstract}

We present an optical through sub-millimetre multi-wavelength study of two very low-luminosity Class I/Flat systems, Mayrit 1701117 and Mayrit 1082188, in the $\sigma$ Orionis cluster. We performed moderate resolution ($R \sim$ 1000) optical ($\sim$0.4--0.9\,$\mu$m) spectroscopy with the TWIN spectrograph at the Calar Alto 3.5-m telescope. The spectra for both sources show prominent emission in accretion- and outflow-associated lines. The mean accretion rate measured from multiple line diagnostics is 6.4$\times$10$^{-10}$ $M_{\sun}$ yr$^{-1}$ for Mayrit 1701117, and 2.5$\times$10$^{-10}$ $M_{\sun}$ yr$^{-1}$ for Mayrit 1082188. The outflow mass loss rates for the two systems are similar and estimated to be $\sim$1$\times$10$^{-9}$\,$M_{\sun}$ yr$^{-1}$. The activity rates are within the range observed for low-mass Class I protostars. We obtained sub-millimetre continuum observations with the Submillimetre Common-User Bolometer Array (SCUBA-2) bolometer at the James Clerk Maxwell Telescope. Both objects are detected at a $\geq$5-$\sigma$ level in the SCUBA-2 850\,$\mu$m  band. The bolometric luminosity of the targets as measured from the observed spectral energy distribution over $\sim$0.8--850\,$\mu$m is 0.18$\pm$0.04 $L_{\sun}$ for Mayrit 1701117, and 0.16$\pm$0.03 $L_{\sun}$ for Mayrit 1082188, and is in the very low-mass range. The total dust+gas mass derived from sub-millimetre fluxes is $\sim$36\,$M_{\rm Jup}$ and $\sim$22\,$M_{\rm Jup}$ for Mayrit 1701117 and Mayrit 1082188, respectively. There is the possibility that some of the envelope material might be dissipated by the strong outflows driven by these sources, resulting in a final mass of the system close to or below the sub-stellar limit.

\end{abstract}

\begin{keywords}
Stars: evolution, protostars -- Submillimetre: stars -- circumstellar matter -- open clusters and associations: individuals ($\sigma$ Orionis)
\end{keywords}

\section{Introduction}
\label{intro}

In the evolutionary scheme defined for a low-mass young stellar object (YSO), the earliest Class 0 stage is characterized by the central object being in a deeply embedded cloud core with the spectral energy distribution (SED) resembling that of a cold blackbody (e.g., Adams et~al. 1987; Lada \& Wilking 1984; Andr\'{e} et~al. 1993). The less embedded stage is Class I, and is characterized by the central object surrounded by a cold circumstellar envelope, an accretion disc, along with outflow/jet/wind activity. In the more advanced Class II stage, the envelope material has completely dissipated, and the star is surrounded by an accretion disc only, with the presence of micro-jets (e.g., Ray et~al. 2007). Finally, for Class III objects, the SED resembles a pure stellar photosphere with the possible presence of a remnant disc. 

Over the past decade, observational surveys conducted in various star-forming regions have revealed a large population of very low-mass stars and brown dwarfs (e.g., Luhman 2012, and references therein). These objects have characteristics, such as, the morphologies and structures of the discs, the dust grain chemical composition, the accretion and jet/outflow properties, the relative disc fractions and lifetimes, which are very similar to Class II and Class III low-mass stars, albeit with the disc masses and accretion rates scaled down according to the mass of the central object (e.g., Muzerolle et~al. 2003; Riaz 2009; Whelan et~al. 2009; Riaz et~al. 2012; Harvey et~al. 2012; Luhman 2012). Such similarities suggest similar evolutionary trends for low-mass stars and very-low mass/sub-stellar objects during these advanced stages. 

There are, however, very few discoveries reported to date on the early-stage Class 0/I objects at the very low-mass end, with $L_{\rm bol}$ $\lesssim$ 0.1 $L_{\sun}$. Surveys in the Taurus star-forming region by e.g., Kenyon \& Hartmann (1995) and White \& Hillenbrand (2004) identified some very-low luminosity Class I objects named IRAS 04158+2805, IRAS 04248+2612, and IRAS 04489+3042, with $L_{\rm bol}$ of $\sim$0.1--0.2\,$L_{\sun}$. A multi-wavelength search in the B213-L1495 clouds in Taurus by Barrado et~al. (2009) and Palau et~al. (2012) led to the discovery of two Class I brown dwarfs or proto-brown dwarfs with total envelope+disc dust mass of the systems, as derived from the sub-millimetre/millimetre fluxes, between 0.3--3 $M_{\rm Jup}$ and 2--20 $M_{\rm Jup}$. Palau et~al. also reported the detection of a candidate pre-sub-stellar core, with an estimated total mass of $\sim$75\,$M_{\rm Jup}$. Pre-sub-stellar cores are considered to be the Class 0 stage or the predecessors of proto-brown dwarfs; these are starless cores with high central densities ($\geq$10$^{5}$ cm$^{-3}$), and show signs of evolving towards forming a proto-brown dwarf (e.g., Tafalla et~al. 1998). Other discoveries of pre-sub-stellar cores include Oph-B11, with an estimated mass of $\sim$20--30\,$M_{\rm Jup}$, in the Ophiuchus region (Andr\'{e} et~al. 2012), and IC348-SMM2E, with $L_{\rm bol}$$\sim$0.1\,$L_{\sun}$, in the IC 348 cluster (Palau et~al. 2014). Likewise, several very low-luminosity objects (VeLLOs) have been discovered in {\it Spitzer Space Telescope} surveys of various star-forming regions (e.g., Di Francesco et~al. 2008; Dunham et~al. 2008). VeLLOs are also high density cores with $L_{int}$$\leq$ 0.1 $L_{\sun}$, where $L_{int}$ is the internal luminosity of the central source itself without any contribution from the surrounding envelope. As with the pre-sub-stellar cores, VeLLOs are also considered to be in the Class 0 stage, and with their very low internal luminosities, it is suggested that these will probably reach a final mass close to or even below the hydrogen-burning mass limit (e.g., Lee et~al. 2009; Dunham et~al. 2008; Takahashi et~al. 2013).

We conducted a search for very low-mass/sub-stellar objects in early Class I stage in the $\sigma$~Orionis cluster. The $\sigma$ Orionis cluster, located around the O9.5-type multiple star of the same name, belongs to the Orion OB 1b association. The X-ray detection of a high concentration of sources around the central star by {\em ROSAT} (Walter et~al. 1994) triggered deep optical surveys dedicated to the search for low-mass stars and brown dwarfs. The cluster has a most probable age of 3$\pm$2 Myr (e.g., Zapatero Osorio et~al. 2002; Sherry et~al. 2004; Caballero 2007) and is located at 380-385 pc (e.g., Caballero 2008; Sim\'{o}n-Diaz et~al. 2011). Deep optical surveys of a large area of the cluster complemented by near-infrared photometry revealed numerous low-mass stars, brown dwarfs (e.g., B\'ejar et~al. 1999), and planetary-mass members (e.g., Zapatero Osorio et~al. 2000). Many objects have been spectroscopically confirmed over a large mass range in the optical (e.g., B\'ejar et~al. 1999; Zapatero Osorio et~al. 2002; B\'ejar et~al. 2001; Barrado y Navascu\'es et~al. 2001; Kenyon et~al. 2005; Caballero et~al. 2006; 2008; Sacco et~al. 2008; Hern\'andez et~al. 2014) and in the near-infrared (e.g., Zapatero Osorio et~al. 2000; Mart\'{i}n et~al. 2001). Follow-up surveys with the {\it Spitzer}, {\em WISE}, and {\em Herschel} space telescopes have revealed a large population of Class II disc systems among these sources, extending into the very low-mass/sub-stellar regime, as well as classical T Tauri analog brown dwarfs with strong accretion activity (e.g., Caballero et~al. 2007; Zapatero Osorio et~al. 2007; Luhman et~al. 2008; Hern\'andez et~al. 2007; 2014; Rigliaco et~al. 2012; Pe\~{n}a-Ram\'irez et~al. 2012; Harvey et~al. 2012). 

This work presents a detailed multi-wavelength characterization of two very low-mass Class I/Flat systems, Mayrit 1701117 (hereafter, M1701117) and Mayrit 1082188 (hereafter, M1082188), identified in the $\sigma$~Orionis cluster. Section \S\ref{obs} describes the near- and mid-infrared observed properties for the targets, and the sub-millimetre and optical observations. Results from radiative transfer modeling of the SEDs, and an analysis of the accretion and outflow activity are presented in Section \S\ref{resultsall}. A discussion on the nature of these sources is presented in Section \S\ref{discuss}.

\section{Targets, Observations and Data Reduction}
\label{obs}

\subsection{Infrared Photometry}
\label{target}

The most complete study in $\sigma$ Orionis comes from the Mayrit catalogue (Caballero 2008) and the UKIDSS Galactic Clusters Survey (Lodieu et~al. 2009), and provides a full census of stars and brown dwarfs down to the deuterium-burning limit over the central 30\,arcmin of the cluster. The sources in these catalogs were searched for counterparts within 10\,arcsec in the Wide Field Survey Explorer ({\em WISE}; Wright et~al. 2010) All-Sky source catalog. We then classified the cross-matched objects by using the traditional method of determining the evolutionary class of a YSO, based on the near- to mid-infrared spectral index, $\alpha_{IR}$ = {\it d} log($\lambda F_{\lambda}$)/{\it d} log($\lambda$) (Lada \& Wilking 1984; Greene et~al. 1994). We used the 2$-$22\,$\mu$m spectral index, and the thresholds of $\alpha_{2-22}$ $>$ 0.3 for Class I sources, $-$0.3 $<$ $\alpha_{IR}$ $<$ 0.3 for Flat Spectrum sources, $-$2 $<$ $\alpha_{IR}$ $<$ $-$0.3 for Class II sources, and $\alpha_{IR}$ $<$ $-$2 for Class III objects. The Class Flat sources are considered to be at an intermediate stage between Class I and II and have tenuous envelopes compared to Class I objects (Greene et~al. 1994). Among the Class I/Flat systems, we found four objects that were relatively faint in the near-infrared ($J >$14\,mag), suggesting that these may be very low-mass/sub-stellar objects. The optical spectra for two of these sources showed a profile similar to AGNs and were therefore discarded. The other two are M1701117 and M1082188, which are characterized in the present work. Both sources were first discovered and catalogued by Caballero (2008). The {\em WISE} matches for both targets are at $\sim$0.05\,arcsec from the target position, with a detection at a signal-to-noise ratio (SNR) of $>$20 in all four {\em WISE} bands. The $\alpha_{2-22}$ index for M1701117 and M1082188 is +0.67$\pm$0.02 and $-$0.23$\pm$0.02, respectively. This index would classify M1701117 as a Class I system, and M1082188 as a Flat Spectrum source. The UKIDSS and {\em WISE} photometry for both targets is listed in Table~\ref{phot}. 

%The typical boundary between low-mass stars and very low-mass objects is considered to be at M $\sim$ 0.4 $M_{\sun}$ (e.g., Chabrier \& Baraffe 1997; 2000). At an age of 3 Myr and a distance of 380 pc for $\sigma$~Orionis, an apparent $J$-band magnitude of 12.5 would imply an absolute $J$ magnitude of 4.6, which would correspond to a mass of $\sim$0.4$M_{\sun}$, using the BT-Settl models (Allard et~al. 2003).

%One of these four objects was found to be the well-published driving source of the Herbig-Haro jet HH 446 (e.g., Andrews et~al. 2004). 

\begin{figure}
     \includegraphics[width=85mm]{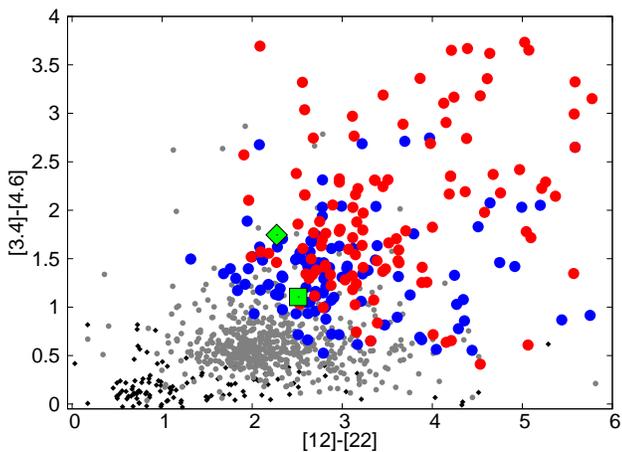} \\        
    \caption{The {\em WISE} mid-infrared colour-colour diagramme with the Class I (red), Flat (blue), Class II (grey), and Class III (black) YSOs from the Evans et~al. (2009) sample. The targets M1701117 and M1082188 are marked as green diamond and square, respectively.    } 
    \label{mid-ir} 
\end{figure}

In Fig.~\ref{mid-ir}, the {\em WISE} mid-infrared colours for M1701117 and M1082188 are compared with the sample of YSOs in five nearby star-forming regions from the work of Evans et~al. (2009). The {\em WISE} counterparts for these YSOs lie within 5\,arcsec of the {\it Spitzer} target position listed in Evans et~al. (2009). Mid-infrared colour-colour diagrammes are particularly useful in separating the Class I protostars from the disc-only Class II objects, and where the Flat sources tend to lie between the Class II and Class I sources. The locations of our targets in the [3.4]$-$[4.6] vs. [12]$-$[22] diagramme in Fig.~\ref{mid-ir} are consistent with the concentrations of Class I/Flat sources, and also indicate M1701117 to be a comparatively less evolved system than M1082188, which lies closer to the Class Flat/Class II boundary. A detailed discussion on the classification of the systems is provided in Section~\S\ref{discuss}. 

%The UKIDSS and WISE photometry for the targets is listed in Table~\ref{phot}. 

\subsection{Sub-millimetre Observations}
\label{sub-mm}

The sub-millimetre observations for the targets were made using the Submillimetre Common-User Bolometer Array (SCUBA-2; Holland et~al. 2013) at the James Clerk Maxwell Telescope. SCUBA-2 is a dual-wavelength (450 and 850\,$\mu$m) camera with 5120 pixels in each of two focal planes. A focal plane consists of four separate sub-arrays, each with 1280 bolometers. The two planes are used simultaneously by means of a dichroic beam-splitter, and have the same field-of-view of $\sim$45\,arcmin on the sky (Holland et~al. 2013). The default map pixels are 2\,arcsec and 4\,arcsec at 450 and 850\,$\mu$m, respectively. The observations were obtained in August, 2013 (PID: M13BU08). Both targets were observed for 170 minutes in Grade 2 weather (225 GHz opacity of 0.06). We used the CV Daisy observing mode, which provides a twice better rms at the centre of the image for small and compact sources of order 3\,arcmin or less, compared to the Pong mode. We also applied a matched-beam filter which utilises the full flux in the beam rather than just the peak value at the position of a source. Using this setup, and with typical grade 2 weather conditions, we expected to reach a 1-$\sigma$ {\it rms} of $\sim$12\,mJy/beam at 450\,$\mu$m, and $\sim$1\,mJy/beam at 850\,$\mu$m.

We used the SMURF (Sub-Millimetre User Reduction Facility), available under the Starlink package, for reducing and calibrating SCUBA-2 data. The various steps required to reduce the data, starting from the pre-processing steps of flat-fielding, etc., to producing the final science map, were applied using the map-making process DIMM (Dynamic Iterative Map-Maker), as described in Chapin et~al. (2013). We used the default DIMM configurations. For post-processing of the science maps, the KAPPA and PICARD softwares were used. For flux calibration, a peak flux conversion factor of 491 Jy/pW/beam and 537 Jy/pW/beam were applied in the 450 and 850\,$\mu$m  bands, respectively, to convert from units of pW to Janskys. The details on calibrating SCUBA-2 data are provided in Dempsey et~al. (2013). We then applied the PICARD recipe SCUBA2\_MATCHED\_FILTER to the flux-calibrated science maps so to improve the point-source detectability. This recipe fits a single Gaussian point spread function (PSF), centred over every pixel in the map, and applies a background suppression filter to remove any residual large-scale noise. The full-width at half maximum of the Gaussian fit was the same as the beam size, which is 7.5\,arcsec and 14.5\,arcsec in the 450 and 850\,$\mu$m  bands, respectively.

\begin{figure*}
     \includegraphics[width=70mm]{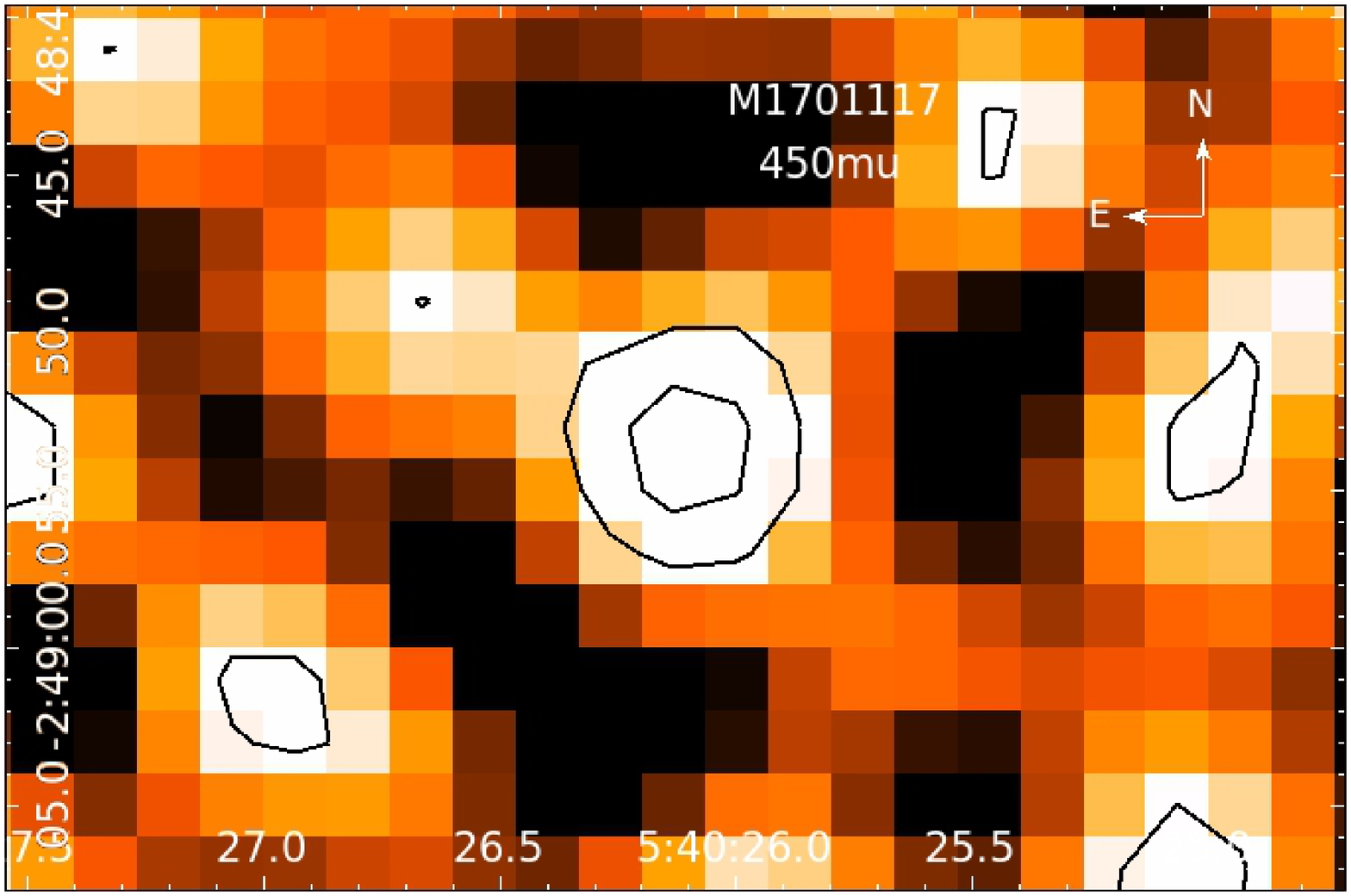} \hspace{0.05in}  \vspace{0.05in}
     \includegraphics[width=70mm]{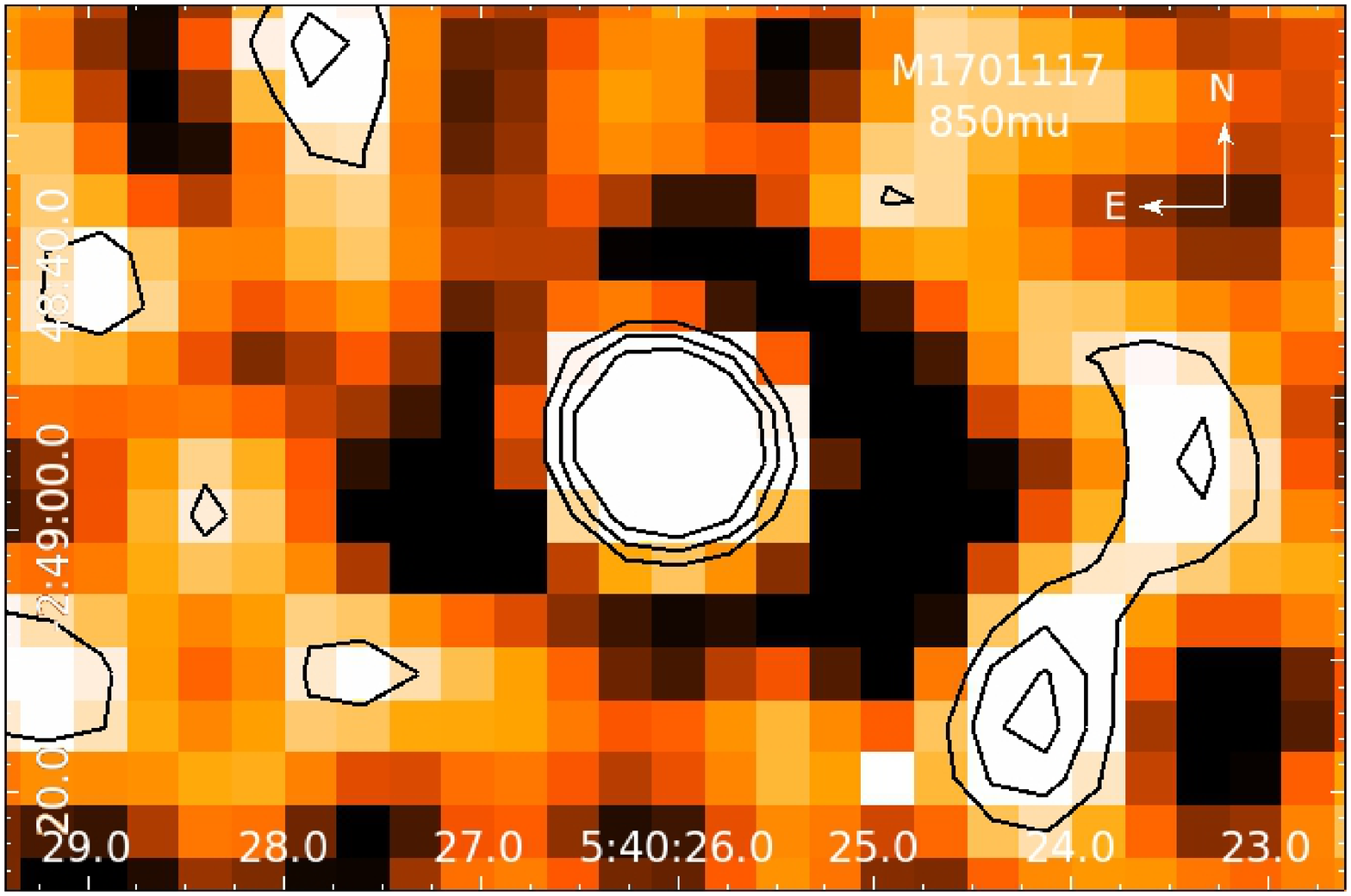} \\ \vspace{0.05in}   
     \includegraphics[width=70mm]{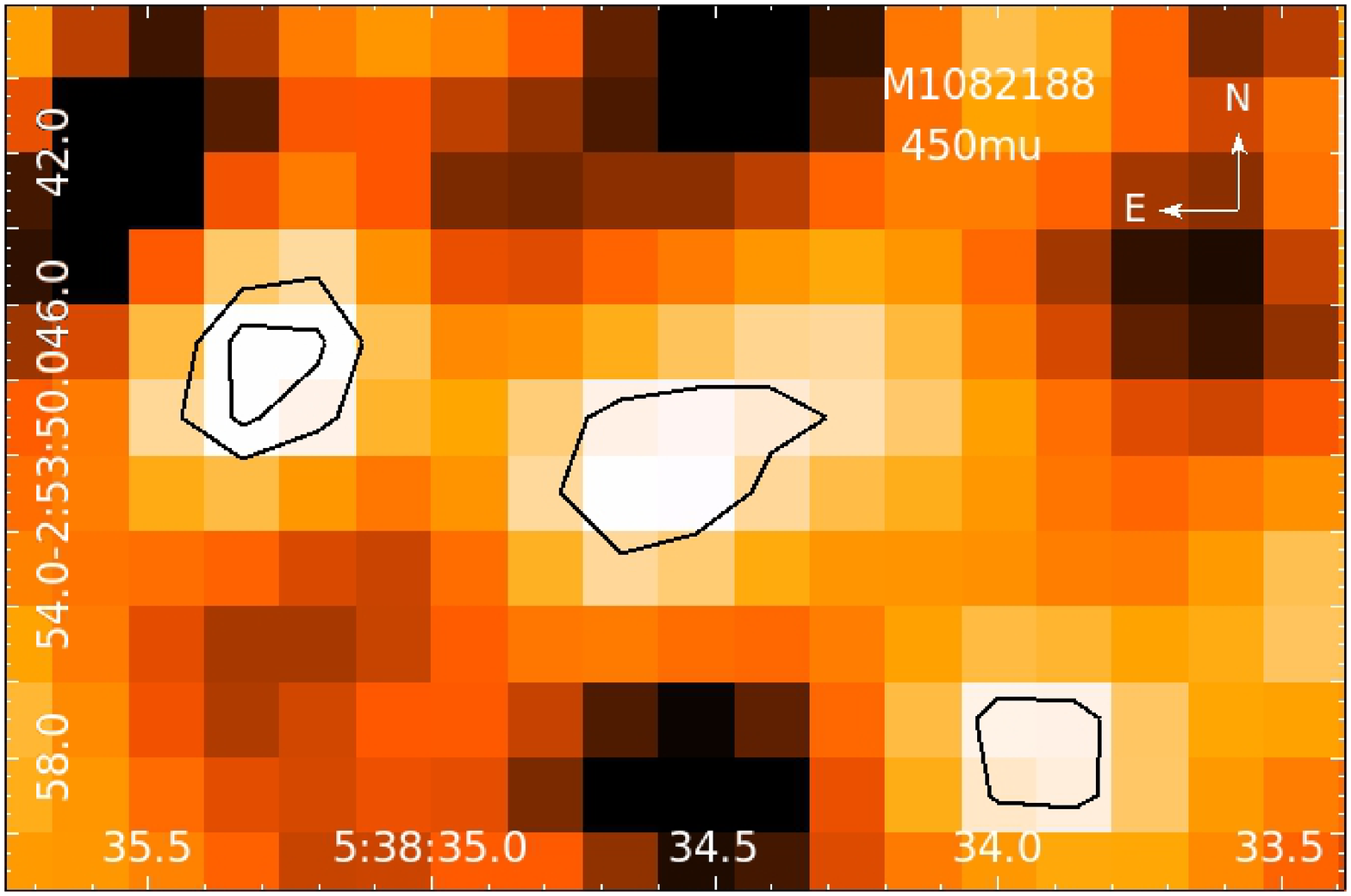} \hspace{0.05in}
     \includegraphics[width=70mm]{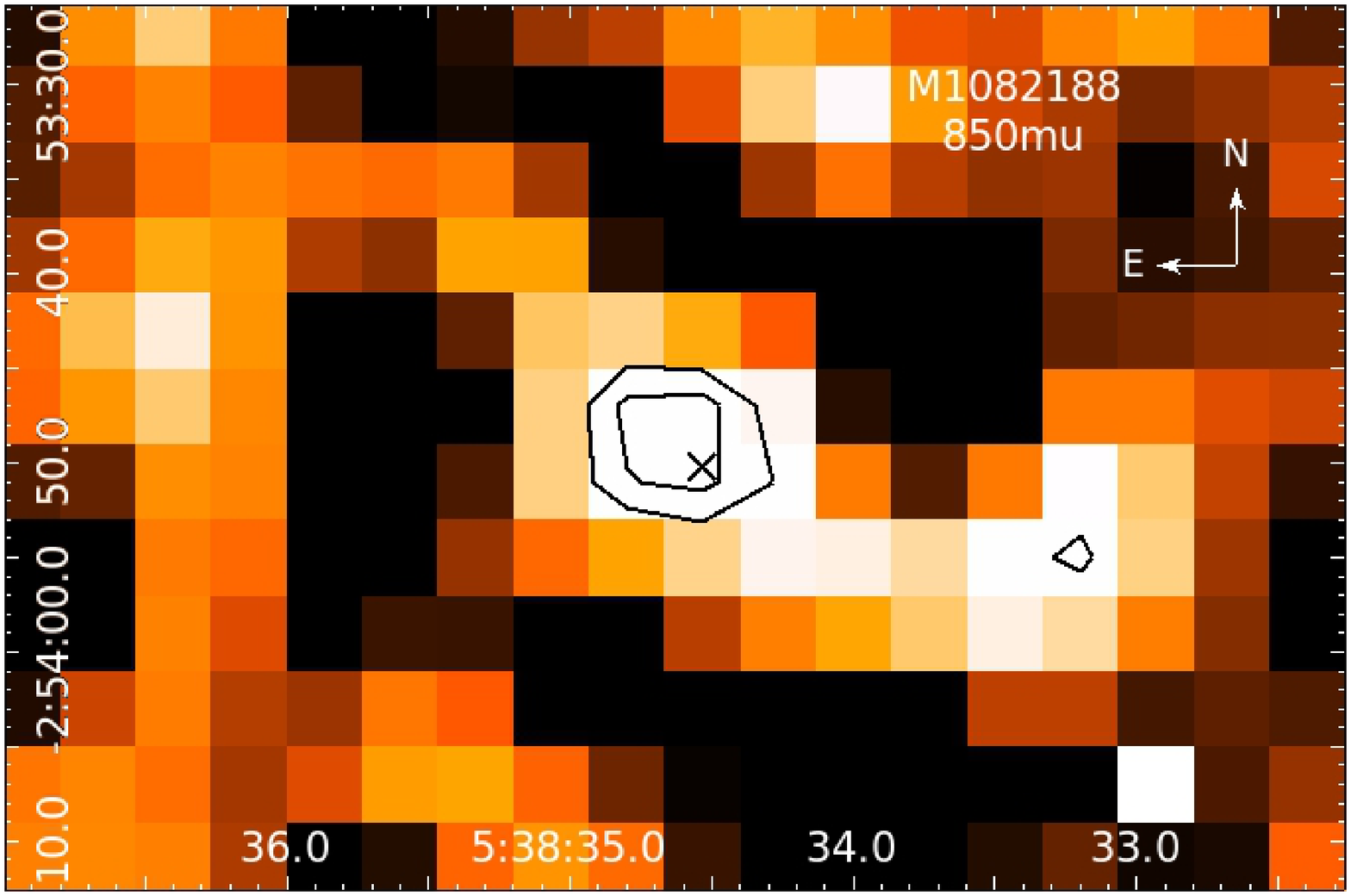} \\         
    \caption{The SCUBA-2 450\,$\mu$m  ({\it left}) and 850\,$\mu$m  ({\it right}) SNR maps for M1701117 ({\it top}) and M1082188 ({\it bottom}). In the 850\,$\mu$m  map for M1082188, the target is marked by a cross. The contours in the 450\,$\mu$m  and 850\,$\mu$m  maps are given from 1$\sigma$ to 3$\sigma$, and 3$\sigma$ to 6$\sigma$, respectively, in steps of 1$\sigma$. North is up, east is to the left. } 
    \label{scuba} 
\end{figure*}

Fig.~\ref{scuba} shows the signal-to-noise maps in the two bands. We used the standard aperture photometry tasks from the IRAF {\it phot} package. The object M1701117 was detected at a $\sim$3-$\sigma$ level in the 450\,$\mu$m  band, and at a $\sim$9-$\sigma$ level in the 850\,$\mu$m  band. The source M1082188 has a fainter $\sim$1.8-$\sigma$ detection in the 450\,$\mu$m  band, but a bright $\sim$5-$\sigma$ level detection in the 850\,$\mu$m  band. The black ring surrounding the 850\,$\mu$m  map of M1701117 is suggestive of an extended source as the PSF filtering does not take into account the error beam. The 850\,$\mu$m  map for M1082188 shows another point source at a distance of $\sim$24\,arcsec south-west from the target location, with possibly a bridge of emission between them. This bridge of emission is at a $<$1-$\sigma$ level, therefore the contribution from it to the PSF of the target is expected to be non-significant. A separation of 24\,arcsec at a distance of 380 pc subtends 9100\,au, which is a distance too large to suggest the presence of extended emission due to a disc or a flattened envelope. Considering that this object shows a peak in emission at a $\sim$2-$\sigma$ level, it is likely to be a spurious or confused source, the proportion of which is found to increase by a factor of four or higher at a SNR threshold of $<$ 4.0 (e.g., Scott et~al. 2002). We calculated the probability of a chance alignment with a background sub-mm galaxy using the SHADES cumulative source counts of Coppin et~al. (2006). For galaxies of the same 850\,$\mu$m flux density as our targets, i.e., 12 and 7.5\,mJy, we expect their on-sky density to be $\sim$50 and $\sim$300 per square degree, leading to a probability of finding a galaxy in a beam-sized aperture centred on the source of 0.07\% and 0.4\%, respectively, which is a negligible probability. The observed sub-millimetre fluxes are listed in Table~\ref{phot}. The 450\,$\mu$m  point for M1082188 should be considered as the 2-$\sigma$ upper limit. 

%The UKIDSS near-infrared images for M1082188 show a point source about 20\,arcsec north-west of the target, which may not be the source seen in the 850\,$\mu$m  map given the different orientation. 

\begin{table}
\begin{minipage}{7cm}
\caption{Photometry}
\label{phot}
\begin{tabular}{ccccc}
\hline
\hline
Band  & M1082188 & M1701117 & Unit & Origin \\ 
\hline
$i'$ & 18.25$\pm$0.008 & 16.54$\pm$0.07 & mag & DENIS \\
$Z$ & 16.594$\pm$0.009 & 16.285$\pm$0.008 & mag & UKIDSS \\
$Y$ & 16.129$\pm$0.008 & 15.973$\pm$0.007 & mag & UKIDSS \\
$J$ & 15.292$\pm$0.006 & 15.455$\pm$0.007 & mag & UKIDSS \\
$H$ & 14.262$\pm$0.005 & 14.435$\pm$0.005 & mag & UKIDSS \\
$K$ & 12.973$\pm$0.002 & 13.333$\pm$0.003 & mag & UKIDSS \\  \relax
[3.6] & 10.990$\pm$0.023 & 11.788$\pm$0.023 & mag & {\em WISE} \\  \relax
[4.5] & 9.943$\pm$0.020 & 10.079$\pm$0.021 & mag & {\em WISE} \\  \relax
[5.8] & 8.117$\pm$0.023 & 6.754$\pm$0.016 & mag & {\em WISE} \\  \relax
[8.0] & 5.711$\pm$0.039 & 4.462$\pm$0.030 & mag & {\em WISE} \\ 
450\,$\mu$m  & $<$40$\pm$20 & 80$\pm$34 & mJy & SCUBA-2 \\
850\,$\mu$m  & 7$\pm$2 & 12$\pm$2 & mJy & SCUBA-2 \\
                     
\hline
\end{tabular}
\end{minipage}
\end{table}

\subsection{Optical Spectroscopy}
\label{optical}

Spectroscopic observations were carried out with the TWIN spectrograph mounted on the Calar Alto 3.5-m telescope in August and December, 2012, in service mode. Weather conditions were photometric and transparency was excellent with a seeing of 1\,arcsec. The TWIN spectrograph is equipped with the Site\#22b (blue) and Site\#20b (red) CCDs. We used the T13 grating in the blue arm to cover the 350--550 nm range, and the T11 grating in the red arm, covering the 550--1100 nm wavelength range. Both are moderate-resolution ($R \sim$1000) gratings. The slit width was set to 1\,arcsec. 

\begin{figure*}
     \includegraphics[width=130mm]{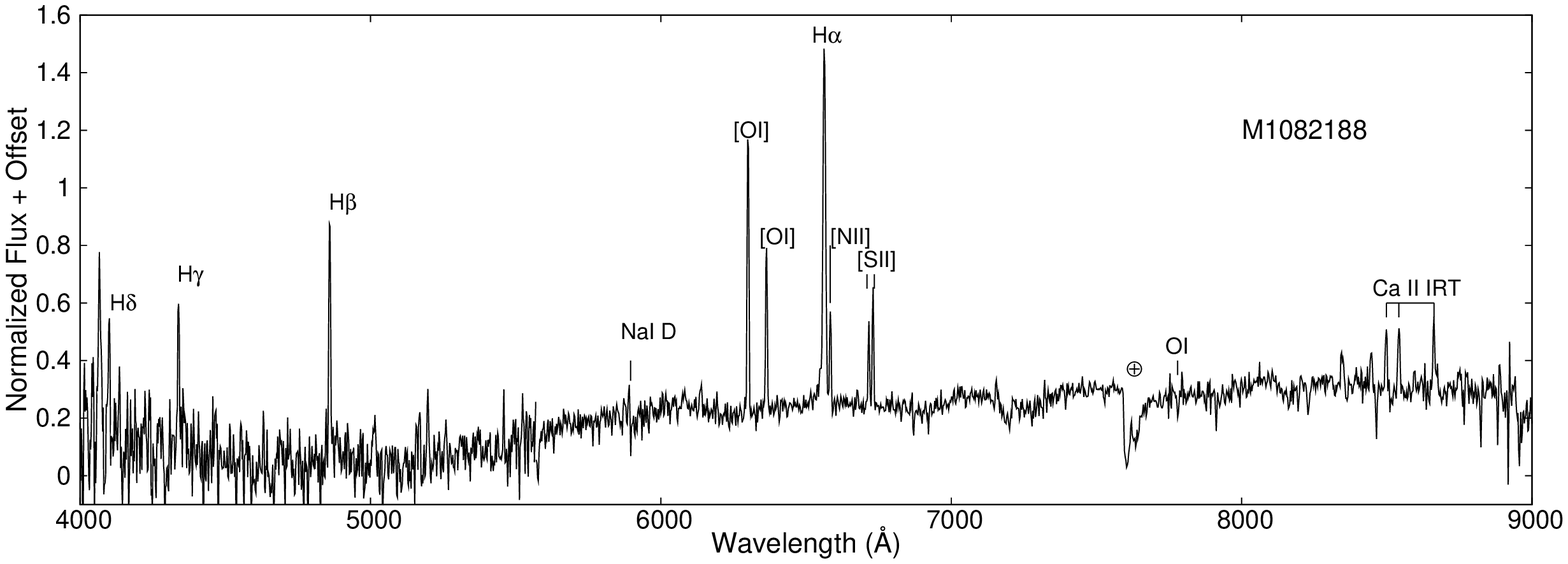} \\   
     \includegraphics[width=130mm]{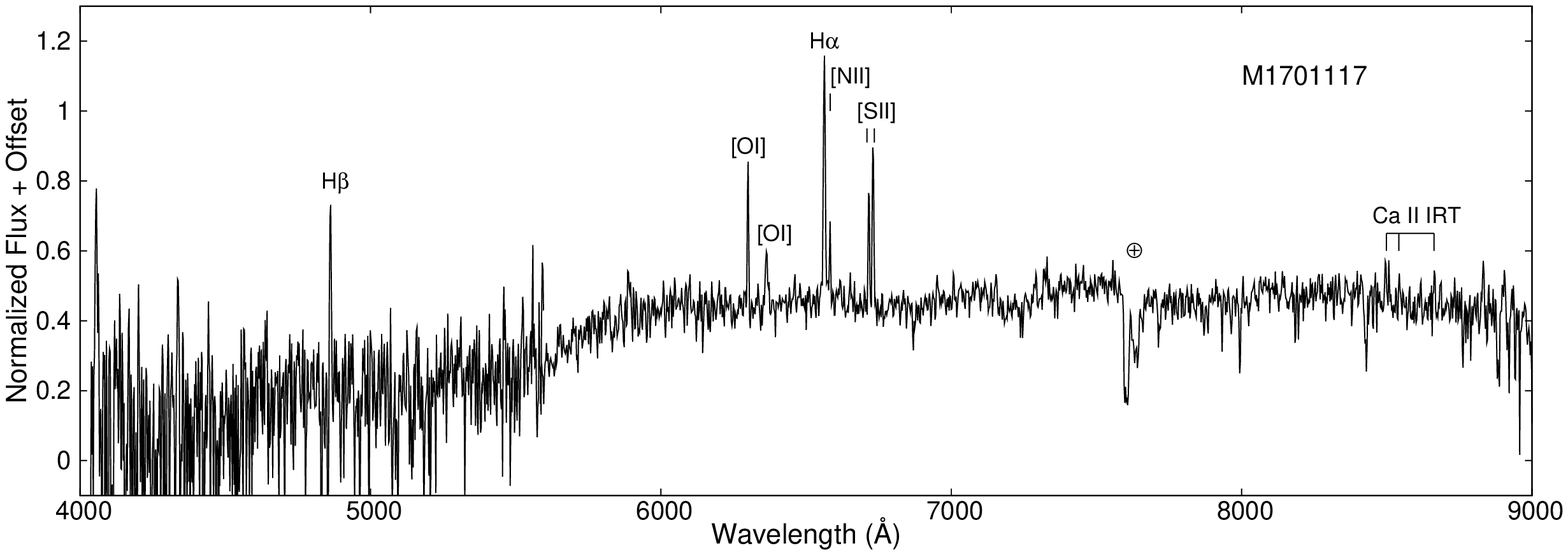} \\    
    \caption{Optical spectrum for M1082188 ({\it top}) and M1701117 ({\it bottom}) with the prominent accretion- and outflow-associated emission lines marked.   } 
    \label{spec-full} 
 \end{figure*}

The exposure time was 1200\,s for the targets in both gratings. The optical spectra were reduced in a standard manner using IRAF routines. We subtracted the bias and divided by the normalised internal flat taken just after the exposures. Then, we extracted optimally the one-dimensional spectrum and calibrated our spectra in wavelength with the helium-argon lamp spectra, to an accuracy better than 0.1\,\AA{}. The spectra were calibrated relative to a spectro-photometric standard (HZ44; Oke 1990) observed as part of our program, and the wavelength scale was corrected to the heliocentric standard of rest. The flux calibration is only valid between $\sim$4000 and 9000\,\AA{}, where flux is well characterized for the spectro-photometric standard. The spectra have not been corrected for telluric absorption. We estimate a signal-to-noise ratio (SNR) of $\sim$5--10 and $\sim$10--20 for the spectra in the blue and red arms, respectively (Fig.~\ref{spec-full}). 

\section{Results}
\label{resultsall}

\subsection{Radiative Transfer Modeling of the SEDs} 
\label{model} 

We used the two-dimensional radiative transfer code by Whitney et~al. (2003) to model the systems. The main ingredients of the model are a rotationally flattened infalling envelope, bipolar cavities, and a flared accretion disc in hydrostatic equilibrium. For the circumstellar envelope, the angle-averaged density distribution varies roughly as $\rho \propto r^{-1/2}$ for $r \ll R_{c}$, and $\rho \propto r^{-3/2}$ for $r \gg R_{c}$. Here, $R_{c}$ is the centrifugal radius and is set equal to the disc outer radius. The disc density is proportional to $\varpi^{-\alpha}$, where $\varpi$ is the radial coordinate in the disc midplane, and $\alpha$ is the radial density exponent. The disc scale height increases with radius, $h=h_{0}(\varpi / R_{*})^{\beta}$, where $h_{0}$ is the scale height at $R_{*}$ and $\beta$ is the flaring power. The disc extends from the dust destruction or the dust sublimation radius, $R_{sub}$ = $R_{*} (T_{sub}/T_{*})^{-2.1}$, to some outer disc radius, $R_{d,max}$. The dust sublimation temperature is adopted to be 1600 K. We used large grains in the dense disc midplane, with a size distribution that decays exponentially for sizes larger than 50\,$\mu$m up to 1 mm. We placed ISM-like grains with $a_{max} \sim$ 0.25\,$\mu$m in the disc atmosphere and the outflow region. The grain model used in the envelope region is similar in size to the ISM-grains, except includes a layer of water ice on the grains that covers the outer 5\% of the radius. Due to binning of photons in the models, there are a total of 10 viewing angles, with face-on covering 0$-$18\,deg inclinations. Bipolar outflow cavities are also included in the models. The cavities extend from the centre of the protostar to the outer radius of the envelope. We adopted the curved cavity shape, the structure of which follows $z = a\varpi^{\beta}$, where $\varpi = (x^{2} + y^{2})^{1/2}$. Here {\it a} is a constant determined by a relation between the envelope radius and the cavity opening angle, and {\it b} is the power of the polynomial defining the cavity shape. The shape parameter determines how quickly the cavity widens in the envelope. A small amount of dust is included in the cavity with constant density, $n_{H_{2}} = 2 \times 10^{4}  cm^{-3}$ (Whitney et~al. 2003). 

%A detailed discussion on the variations in the model SEDs with envelope and disc parameters can be found in Riaz et~al. (2009) and Riaz \& Gizis (2007).

\begin{figure}
     \includegraphics[width=80mm]{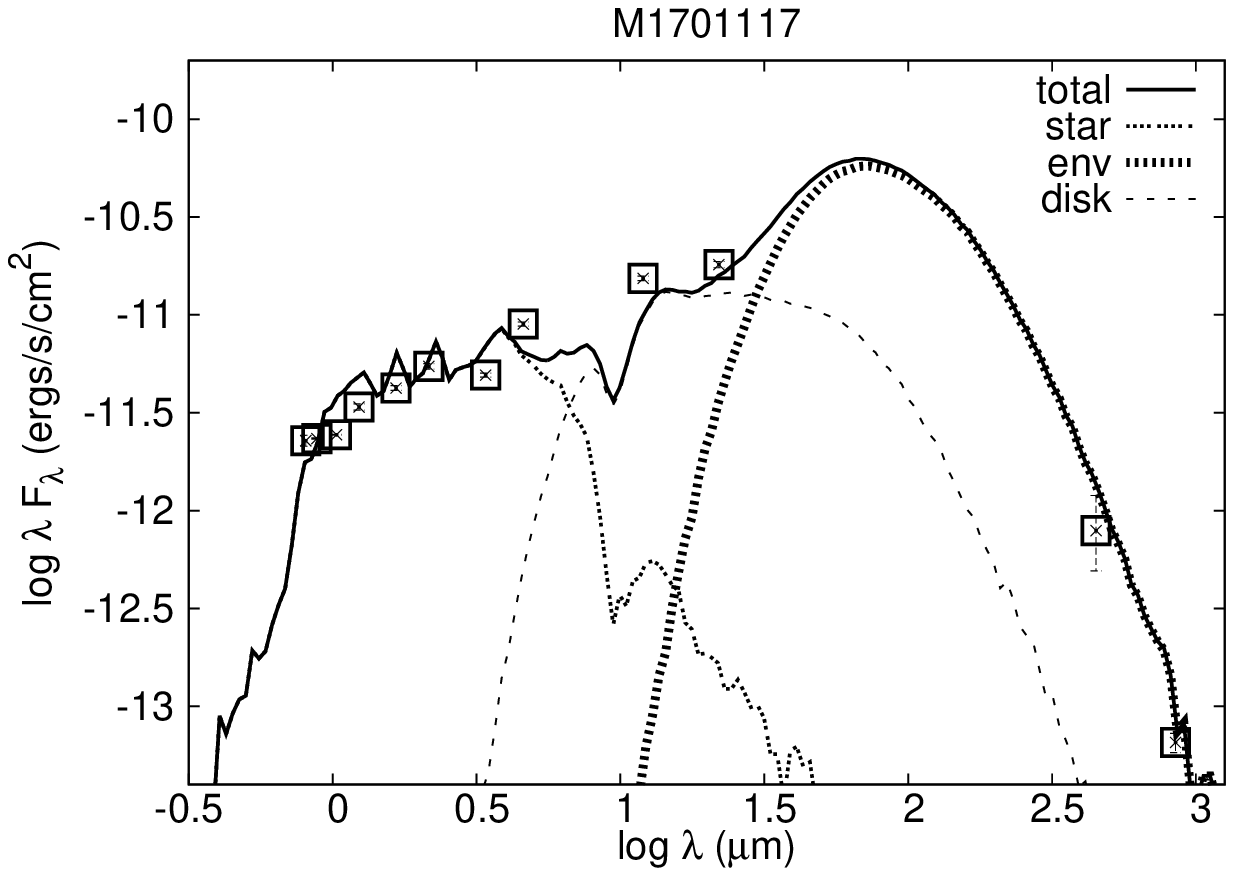}  
     \includegraphics[width=80mm]{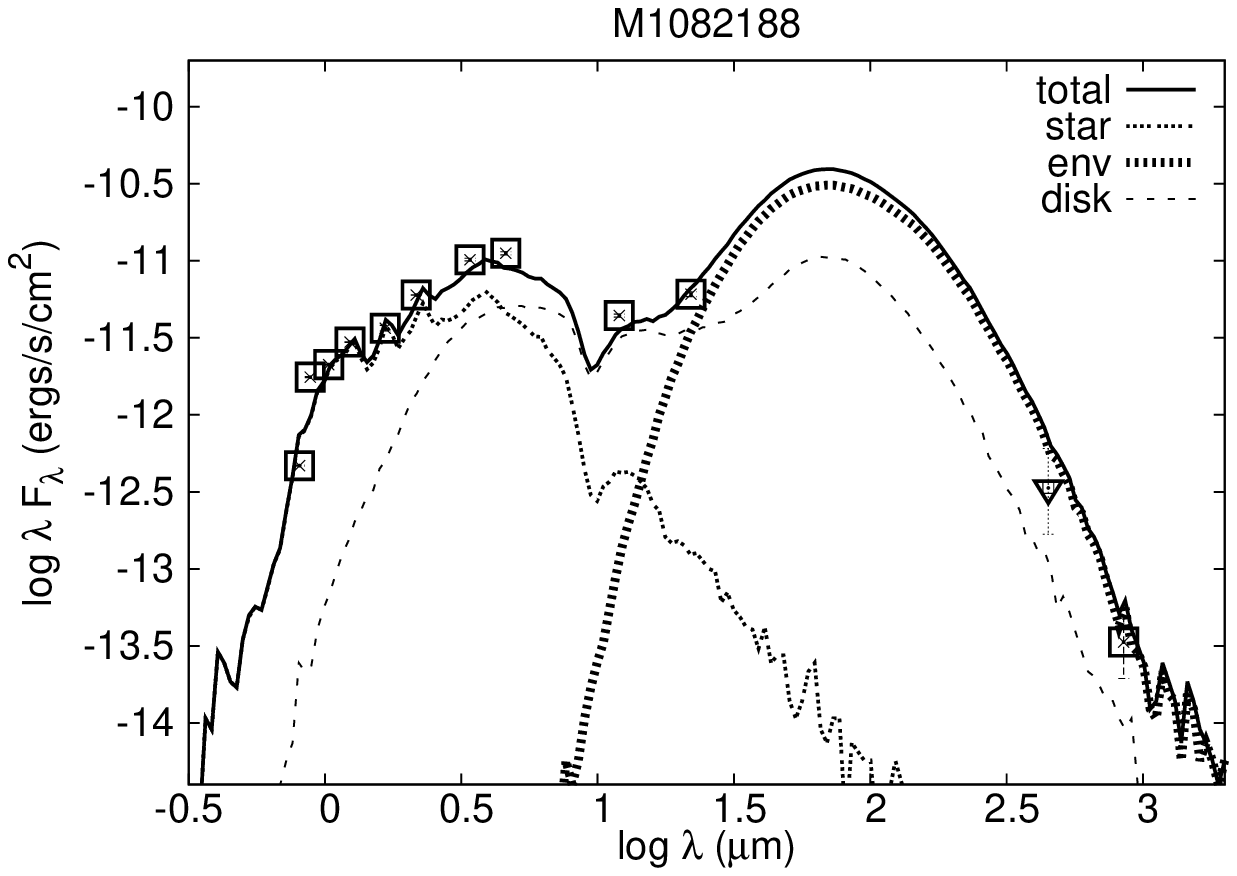}  
    \caption{The best-fit from radiative transfer modeling of the SEDs for the M1701117 ({\it top}) and M1082188 ({\it bottom}) systems. Also shown are separate contributions from the star, disc, and the envelope. The optical to sub-millimetre photometry is plotted with open squares.  } 
    \label{modeling} 
 \end{figure}

Table~\ref{modelfit} lists the estimates on the various model parameters; the top row lists the best-fit values based on the lowest $\chi^{2}$ value of the fit to the observed SED, while the bottom row shows the range in values for each parameter based on the degeneracies in the model fits obtained from the top three fits. The best-fits are shown in Fig.~\ref{modeling}. An important parameter in fitting a Class I SED is the mass infall rate, since increased/decreased envelope infall rates correspond to denser/thinner envelopes. A thinner envelope implies that more photons can escape through the cavity regions, resulting in a higher near-infrared flux. Thus both the sub-millimetre and the near-infrared points provide a good constraint to this parameter. The best model fits to the observed SEDs for the targets were obtained using an infall rate of (3--8)$\times$10$^{-6}$ $M_{\sun}$ yr$^{-1}$. Increasing the mass infall rate to values $>$1$\times$10$^{-5}$$M_{\sun}$ yr$^{-1}$ results in a model fit with large excess emission in the sub-millimetre region and misses the 850\,$\mu$m  point. The best model fits shown in Fig.~\ref{modeling} indicate some amount of flaring between $\sim$20\,$\mu$m and the sub-millimetre points. In order to obtain a flat structure in this wavelength region, the infall rate would need to be reduced to 1x10$^{-7}$$M_{\sun}$ yr$^{-1}$ and the envelope mass reduced to just 1 $M_{\rm Jup}$. Such a model, however, falls well short of fitting the sub-millimetre points for both targets. We therefore expect some flaring in the system, and not a completely flat structure. We found a good fit using intermediate inclinations of 45-65\,deg. A smaller inclination angle results in a model with larger near-infrared fluxes and more emission at {\em WISE} wavelengths than the observed fluxes. Among the other parameters, the outer envelope radius mainly effects the SED for $\lambda >$ 100\,$\mu$m, but very large radii can also give too much optical depth to the centre of the envelope, thus effecting the mid-infrared fluxes. This parameter was constrained by the {\em WISE} and sub-millimetre fluxes, and values of around $\sim$1500\,au provide a good fit. 

%Best model fits indicate an extinction from the circumstellar material of $A_{V}$$\sim$9 mag for M1082188 and $\sim$15 mag for M1701117. The latter source has a higher envelope mass than the former, and therefore a higher extinction is expected. This $A_{V}$ estimate is the total extinction measured from the outside of the YSO through the envelope and disc to the stellar surface along the line of sight, and takes into account the column density of the envelope, disc, and bipolar cavity combined with the dust opacity (Whitney et~al. 2003). 

\begin{table}
\caption{Model Parameters}
\label{modelfit}
\begin{tabular}{lcc}
\hline
\hline
Parameter  & M1082188 &  M1701117 \\
\hline 

%$M_{*}$ & 40 $M_{\rm Jup}$ & 40 $M_{\rm Jup}$ \\
%$R_{*}$  & 4.7 $R_{\rm Jup}$ & 4.7 $R_{\rm Jup}$ \\
%$T_{eff}$ & 2800 K & 2800 K \\  \hline

$\dot{M}_{env}$ ($M_{\sun}\,yr^{-1}$) & 3.95$\times$10$^{-6}$ & 7.66$\times$10$^{-6}$ \\
& (3--7)$\times$10$^{-6}$ & (4--8)$\times$10$^{-6}$ \\ 

\smallskip

$M_{\rm env}$ ($M_{\rm Jup}$) & 40 & 50  \\
& 23--55  & 30--65  \\ 

\smallskip

$R_{\rm sub}$ ($R_{*}$) & 3.12  & 3.12  \\ 

\smallskip

$R_{\rm env,min}$ ($R_{sub}$) & 1  & 34.8  \\
& 1 & 20--35  \\ 

\smallskip

$R_{\rm env,max}$ (au) & 1457  & 1765  \\
& 1380--1460  & 1650--1770 \\ 

\smallskip

$M_{\rm disc}$ ($M_{\rm Jup}$) & 1--4  & 1--4  \\ 

\smallskip

$R_{\rm d,min}$ ($R_{sub}$) & 1  & 30  \\
& 1 & 20--35 \\ 

\smallskip

$R_{\rm d,max}$ (au) & 108  & 42  \\
& 30--110  & 37--45  \\ 

\smallskip

$R_{\rm c}$ (au) & 108  & 42  \\
& 30--110 & 37--45  \\ 

\smallskip

$\theta_{\rm in}$ & 45$-$60 deg & 50$-$65 deg \\ 

\smallskip

$\theta_{\rm cav}$ &  21.3 deg & 28.8 deg \\
& 21$-$29 deg & 28$-$29 deg \\ 

\smallskip

$\beta$ & 1.138 & 1.149 \\
& 1.07--1.14 & 1.08--1.15 \\ 

\smallskip

$\alpha$ & 2.138 & 2.149 \\
& 2.07-2.14 & 2.08-2.15 \\ 
   
\hline 
\end{tabular}
\end{table}

The best-fit parameters for both targets are similar to the Class I/Late Class I stage standard models presented in Whitney et~al. (2003). Comparing the two objects, M1701117 is brighter in the sub-millimetre bands than M1082188, and requires a larger envelope mass to fit these points (Fig.~\ref{modeling}; Table~\ref{modelfit}). We also tried to fit the observed SEDs using a disc-only model, with the envelope infall rate set to zero. Disc emission alone cannot fit the full SED, particularly the rise in fluxes between $\sim$2 and 5\,$\mu$m and the sub-millimetre points. We also checked with the online SED fitting tool (Robitaille et~al. 2006)\footnote{\url{http://caravan.astro.wisc.edu/protostars/}}, and all of the `top ten' model fits for both targets include the envelope component. 

%The disc mass from the model fit for M1701117 is lower by a factor of $\sim$10 compared to M1082188, which suggests that the disc in this system is in the initial stages of formation and is another indication of M1701117 being a comparatively less evolved system. 

We derived the total dust+gas mass from the envelope+disc components of the systems from the 850\,$\mu$m flux. This mass estimate is $\sim$36\,$M_{\rm Jup}$ and $\sim$22\,$M_{\rm Jup}$ for M1701117 and M1082188, respectively. These masses have been derived assuming a dust temperature of 10 K, a gas-to-dust mass ratio of 100, a dust opacity law of 0.017 cm$^{2}$ g$^{-1}$ (Ossenkopf \& Henning 1994), and  a distance to $\sigma$~Orionis of 380 pc (e.g., Caballero 2008). The dust temperature was estimated by solving the radiative transfer through the model envelope given the total luminosity of the source, and is consistent with the average isothermal dust temperatures found from radiative transfer models of a sample of Class I stars from the work of e.g., Shirley et~al. (2002) and Young et~al. (2003). These circumstellar mass estimates are consistent with the lower value on the range of the total envelope+disc mass obtained from the SED modeling (Table~\ref{modelfit}).

\subsection{Accretion and Outflow Signatures}
\label{emlines}

The optical spectra for both targets exhibit strong emission in the accretion-associated lines of H$\alpha$ and the Ca~{\sc ii} infrared triplet (IRT) at 8498, 8542, 8662 $\AA$, as well as in the outflow-associated forbidden emission lines (FELs) of [O~{\sc i}] $\lambda$$\lambda$6300, 6363\,{\AA}, [S~{\sc ii}] $\lambda$$\lambda$6716, 6730\,{\AA}, and the [N~{\sc ii}] line at 6583$\AA$ (Fig.~\ref{spec-full}). The presence of such youth signatures is a confirmation of both targets being YSOs, and not foreground/background stars or extragalactic contaminants. In comparison with M1082188, the strength in all emission lines, except the [S~{\sc ii}] FELs, are weaker in the spectrum of M1701117 (Table~\ref{fluxes}). This can be expected since the envelope mass for this system is larger than M1082188, and this system is therefore more extinguished by the cold envelope, making it fainter in the optical. The spectra for both targets are strongly veiled, which makes it difficult to identify any photospheric absorption lines or bands and measure the systemic velocity. We considered the rest velocity, $V_{LSR}$= 30.9\,km s$^{-1}$, for both sources, which is the mean value of the radial velocities measured for several members of the $\sigma$~Orionis cluster by Sacco et~al. (2008). 

The presence of optical FELs is not commonly seen in Class I protostars, due to extinction from the envelope and thus being too faint or undetected in the optical bands. There are, however, a few known cases of Class I systems, such as, GV Tau, HL Tau, IRAS 04369+2539, IRAS 05451+0037, and in particular, the low-luminosity source IRAS 04158+2805 ($L_{bol} \sim$0.3\,$L_{\sun}$), which show both strong emission in the optical FELs as well as some of the other signatures observed in protostars (e.g., White \& Hillenbrand 2004; Hillenbrand et~al. 2012; Furlan et~al. 2008). These objects are treated as diffuse envelope systems, which are likely more evolved and at the end of their Class I phase. The tenuous envelopes of our targets, in addition to outflows, suggests that these may be rare cases among very low-mass mass objects as the few found among protostars.

\begin{figure}
\centering
     \includegraphics[width=40mm]{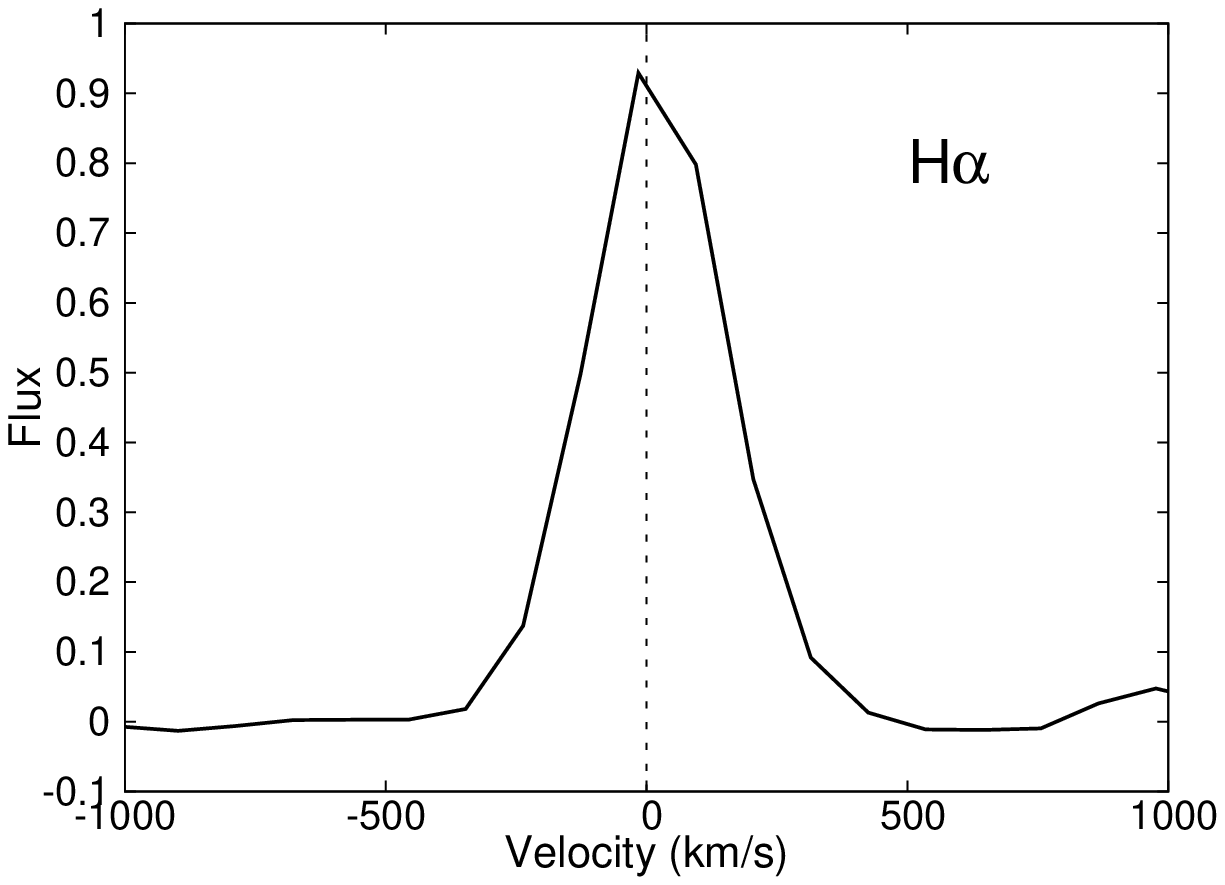} 
     \includegraphics[width=40mm]{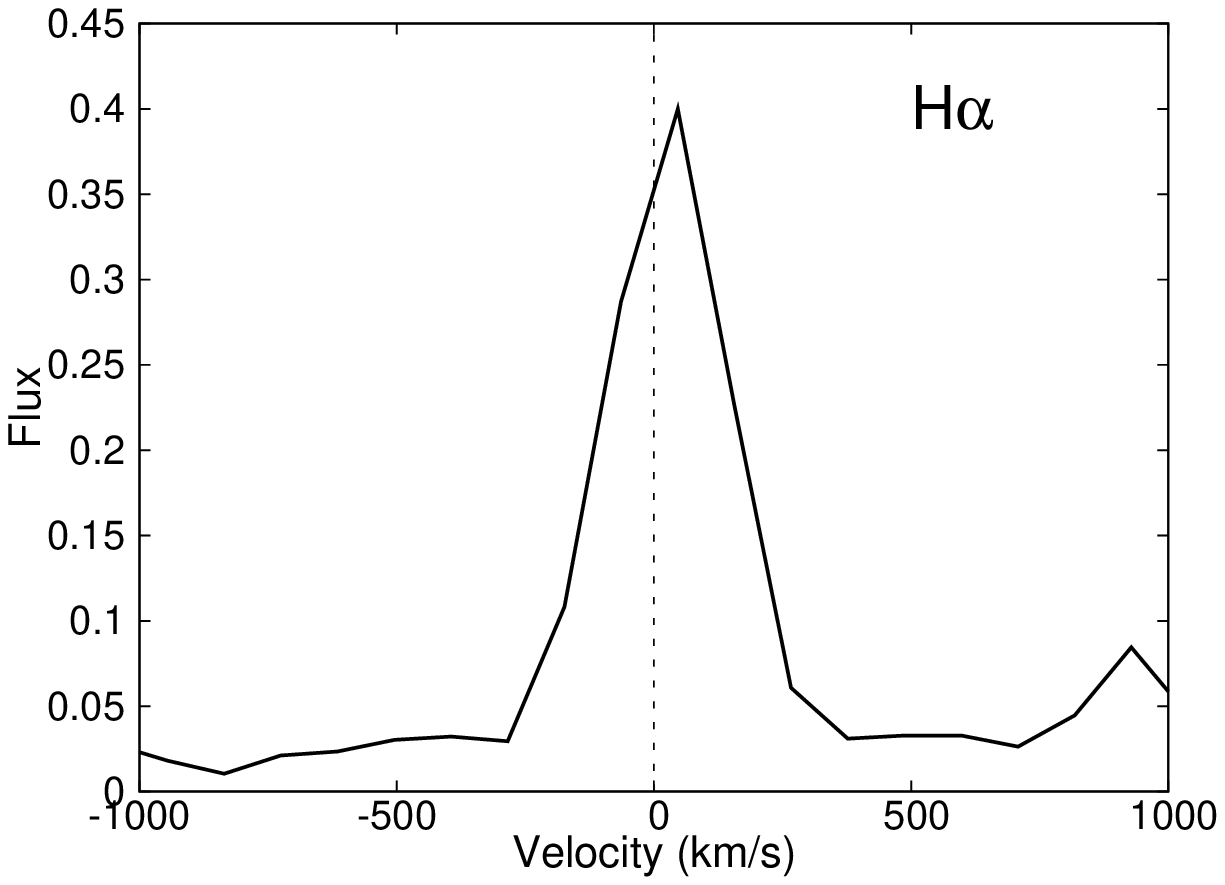} \\

     \includegraphics[width=40mm]{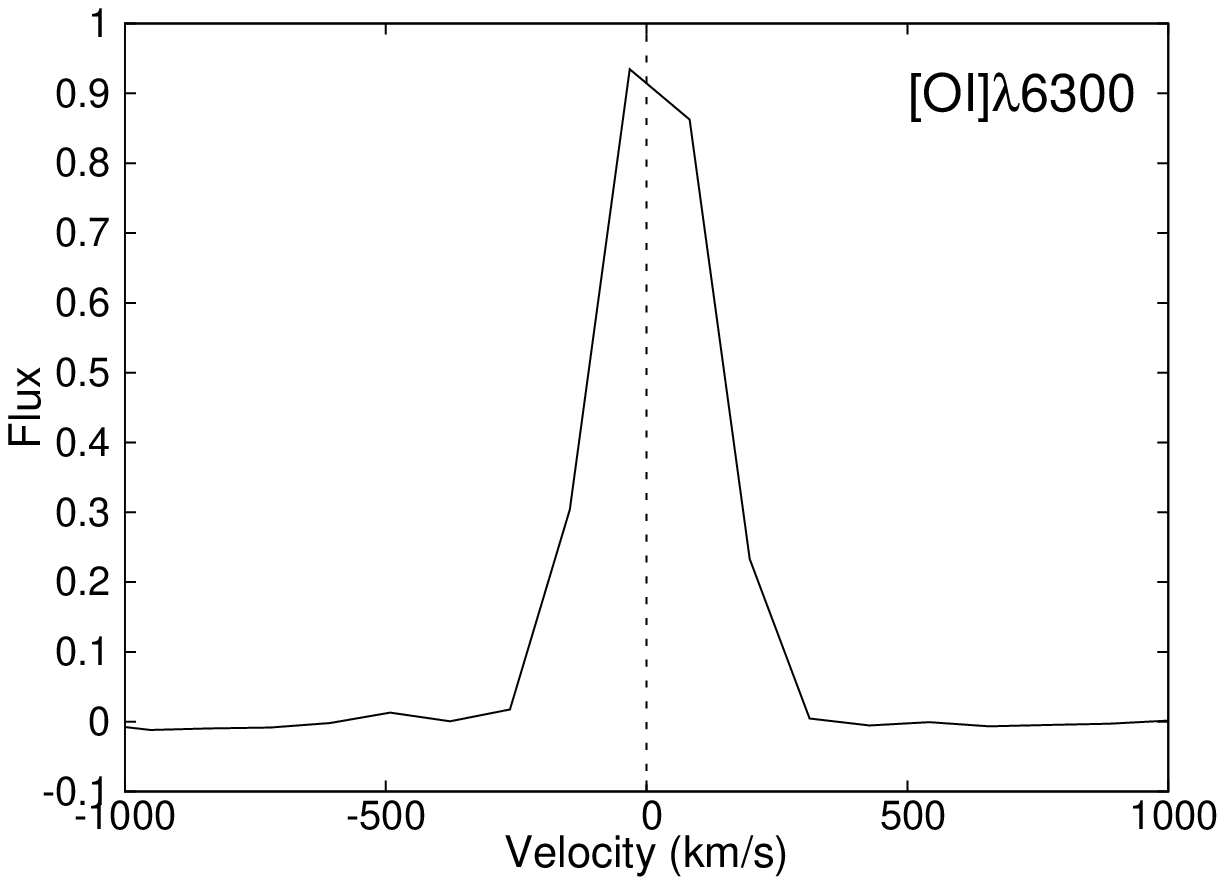}                      
     \includegraphics[width=40mm]{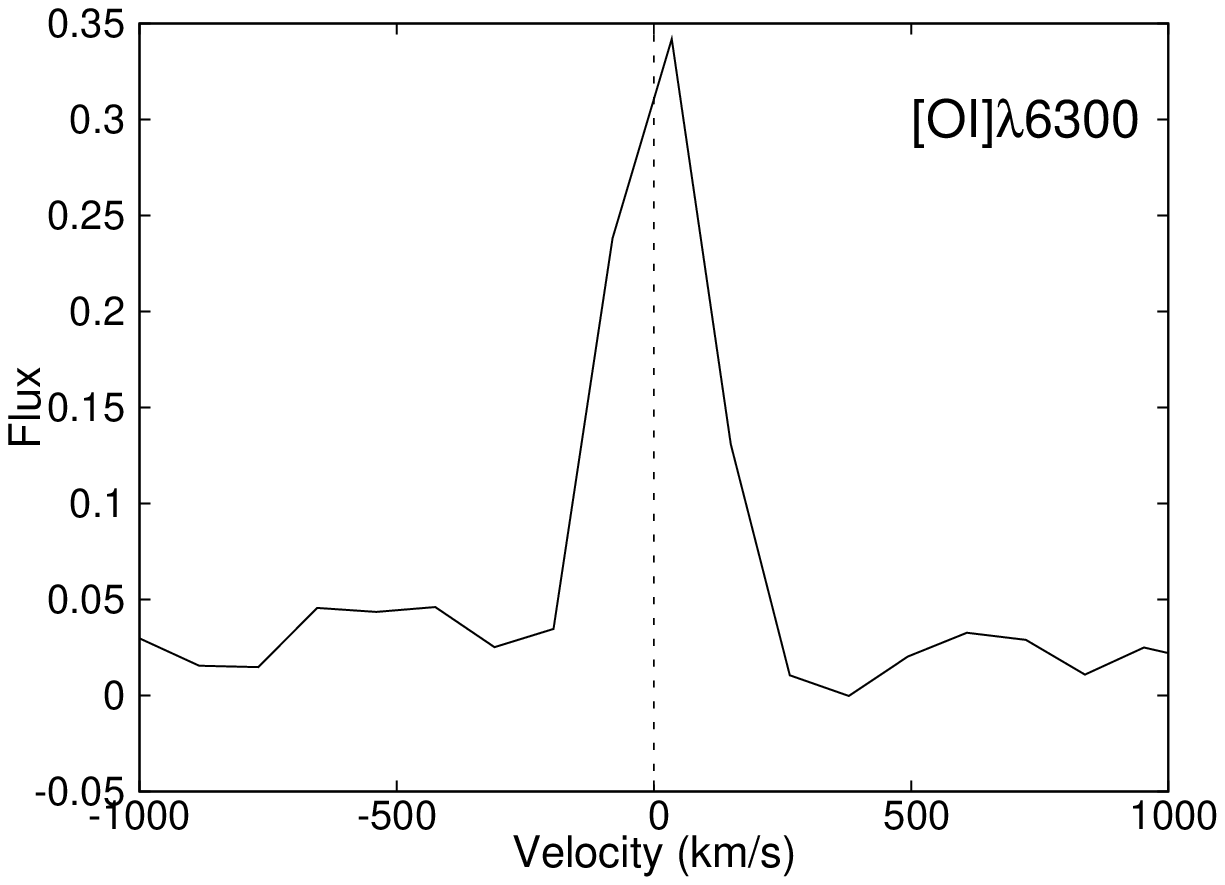}      \\

     \includegraphics[width=40mm]{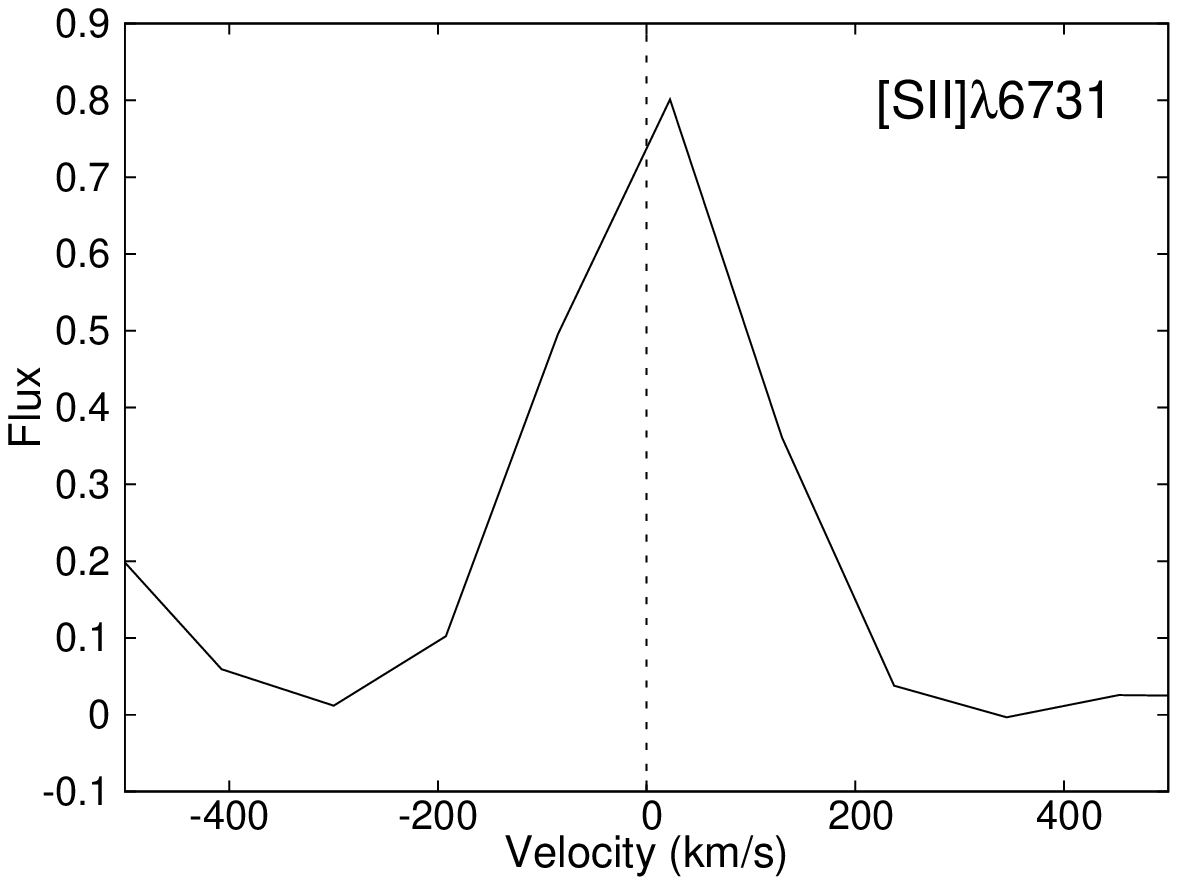}    
     \includegraphics[width=40mm]{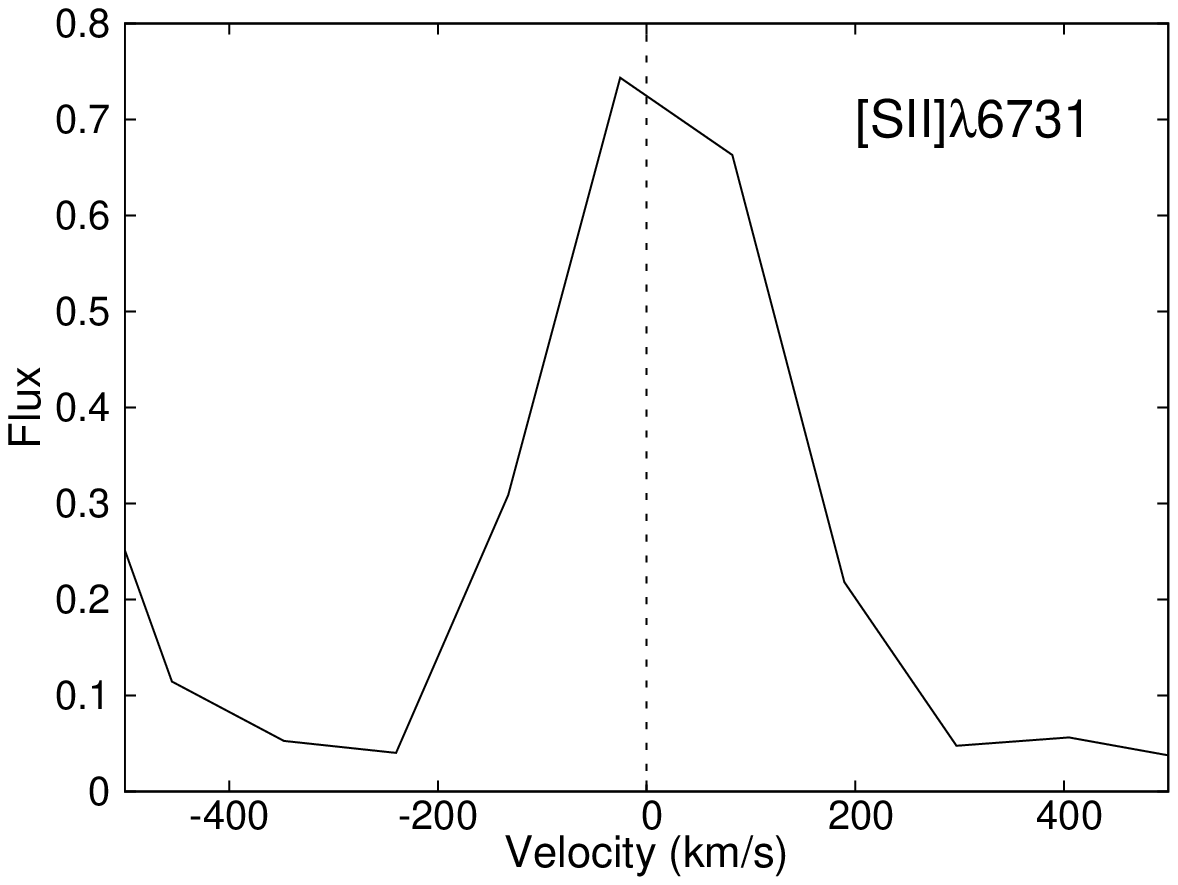}    \\
          
     \includegraphics[width=40mm]{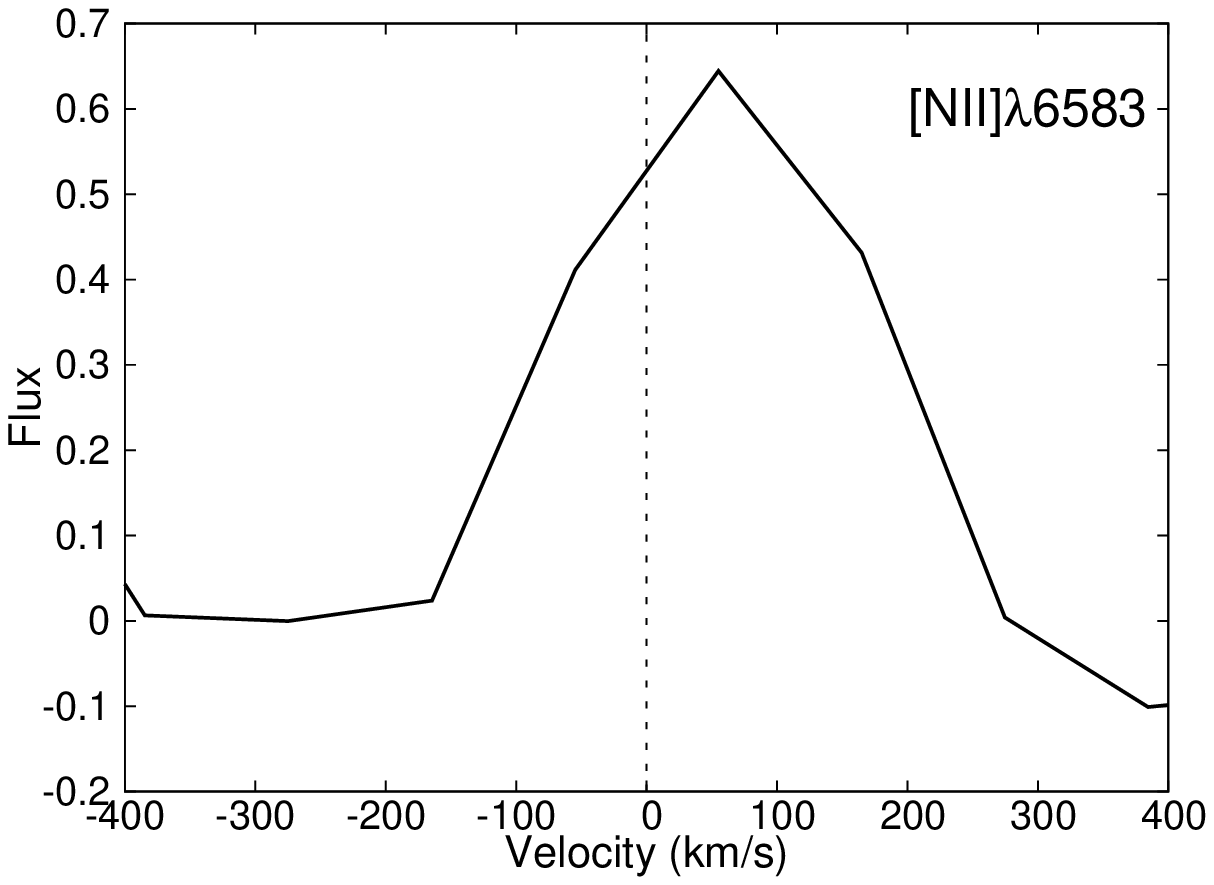}  
     \includegraphics[width=40mm]{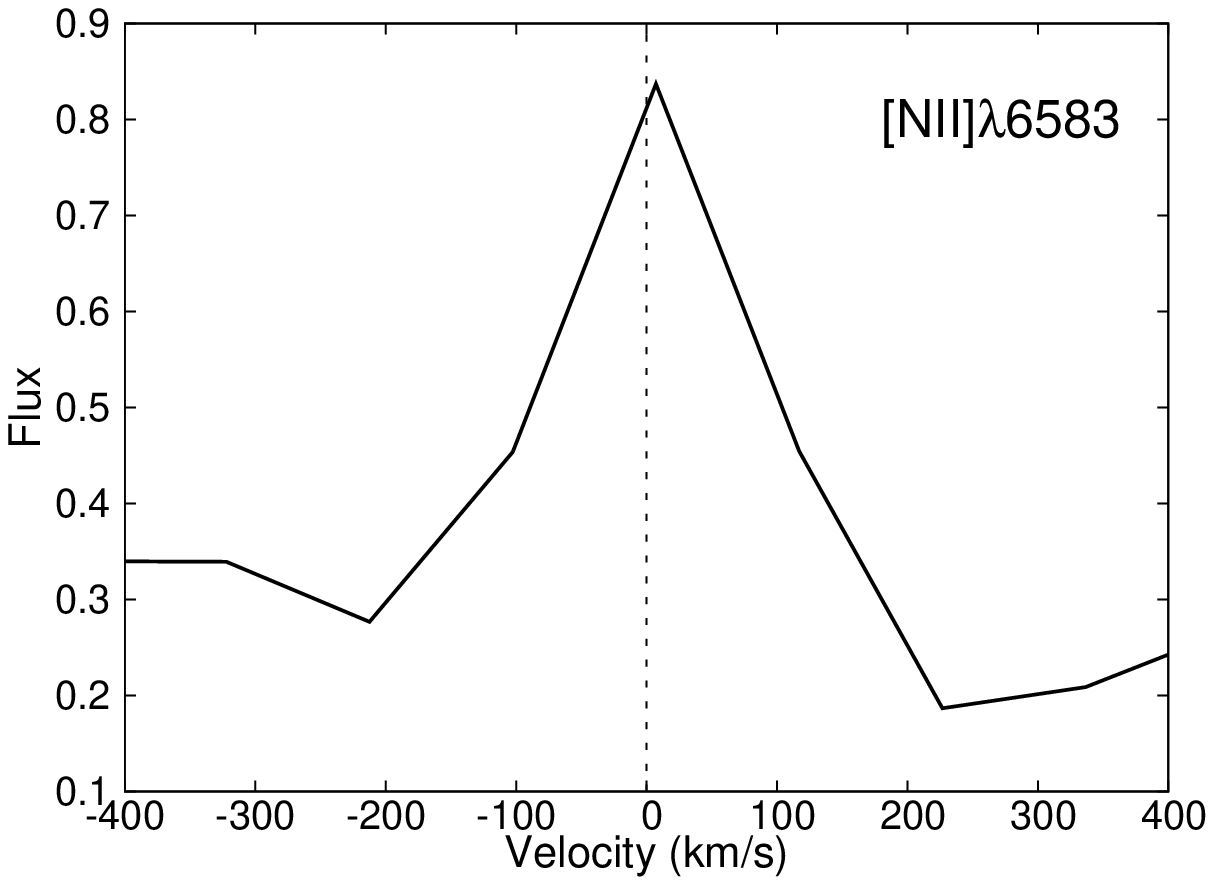}       \\
     
    \caption{Accretion and outflow associated emission lines for M1082188 ({\it left}) and M1701117 ({\it right}). All radial velocities are with respect to the systemic velocity assumed for the targets. The flux is in arbitrary units, roughly normalized to the line peak value. } 
    \label{lines} 
 \end{figure}

For both targets, the H$\alpha$ line shows a broad profile (Fig.~\ref{lines}), with a width at 10\% of the line peak of 580$\pm$60 km s$^{-1}$ and 380$\pm$40 km s$^{-1}$ for M1082188 and M1701117, respectively. An H$\alpha$ 10\% width of $\geq$200 km s$^{-1}$ is usually considered as a threshold to distinguish between accreting and non-accreting sources (e.g., Muzerolle et~al. 2003), based on which both targets can certainly be classified as intense accretors. The spectrum of M1082188 also shows strong emission in Ca~{\sc ii} IRT, which is another important accretion indicator. The strength in the [S~{\sc ii}] FELs is stronger for M1701117. The difference in the ratio of the [S~{\sc ii}] lines between the two sources is likely due to differences in the electron density, n$_{e}$, in the outflows. We used the [S~{\sc ii}] line ratios to calculate n$_{e}$ of 3600$\pm$1000 cm$^{-3}$ and 4575$\pm$50 cm$^{-3}$ for M1082188 and M1701117, respectively. On the blue side of the spectrum of M1082188 (Fig.~\ref{spec-full}), the upper Balmer lines of H$\gamma$ and H$\beta$ are prominently detected, while there is a weak detection for the H$\delta$ line. The H$\beta$ line is also a notable accretion indicator (e.g., Fang et~al. 2011). The blue side of the M1701117 spectrum is at a much worse SNR, with only the H$\beta$ line clearly identified. The strength in this line is also weaker compared to M1082188. 

%In comparison, the Ca~{\sc ii} triplet for M1701117 is weakly detected, and the lines are severely noise-affected. For the case of M1082188, the broadest profile among all FELs is observed for the [NII]$\lambda$6583 line. In contrast, the [NII] FEL for M1701117 is narrowly peaked and centred at the line centre. The [S~{\sc ii}] profiles for M1701117 are broader in shape, unlike the narrow profiles observed for M1082188. 

%The difference in the shapes of the [S~{\sc ii}] lines could be due to the fact that the spectral resolution is low and the lines are poorly sampled. For some reason, the [S~{\sc ii}] emitting region in M1701117 is broader in velocity, but it would be difficult to draw any further conclusions without spectra along the outflow with a better velocity resolution. 

It is important to note that the FELs of both objects are centred at zero velocity. If the radial velocities of these lines were $\sim$100--200\,km s$^{-1}$, as observed for jets from low-mass protostars (e.g., White \& Hillenbrand 2004), we would have seen these shifts even with the poor spectral resolution. The fact that the lines lie close to zero velocity indicates that the radial velocities for these Class I jets are similar to the radial velocities measured for jets in Class II very low-mass/sub-stellar objects, which are typically measured to be $\sim$50\,km s$^{-1}$ (e.g., Whelan et~al. 2009). We discount the idea that the low radial velocities are due to the outflows being in the plane of the sky as there is no evidence that our sources have edge-on discs (Table~\ref{modelfit}). A low jet velocity for Class I/II very low-mass star or brown dwarf is expected, considering that the escape velocity from a brown dwarf should be less than a low mass star, as the escape velocity is related to the stellar mass.

\begin{table*}
\caption{Emission line fluxes and equivalent widths}
\label{fluxes}
\begin{tabular}{cccccccc}
\hline
\hline
Line & $\lambda_{central}$ ($\AA$)   & \multicolumn{2}{c}{Equivalent Width ($\AA$)}  & \multicolumn{2}{c}{Line Flux (10$^{-15}$ erg cm$^{-2}$ s$^{-1}$)}  & \multicolumn{2}{c}{log $\dot{M}_{acc}$ ($M_{\rm \sun}$ yr$^{-1}$) }    \\ 
         &                     & M1082188 & M1701117 &  M1082188 & M1701117  &  M1082188 & M1701117    \\
\hline

H$\delta$           & 4101.74 & $-$36.8$\pm$2.9       & --                          & 1.7$\pm$0.2    &  -- & $-$9.30$\pm$0.42 & -- \\ 
H$\gamma$      & 4340.47 & $-$23.9$\pm$4.1       & --                           & 1.5$\pm$0.2    & --  & $-$9.52$\pm$0.36 & --  \\
H$\beta$            & 4861.33 &  $-$56.9$\pm$4.5      & $-$49.6$\pm$4.9   & 2.1$\pm$0.6   & 2.6$\pm$0.2 & $-$9.44$\pm$0.34 & $-$9.58$\pm$0.34 \\  \relax
[O~{\sc i}]                       & 6300.30 & $-$122.8$\pm$11.0  & $-$14.7$\pm$1.2   & 6.7$\pm$0.5   & 1.8$\pm$0.1 & $-$7.81$\pm$0.41 & $-$8.16$\pm$0.41  \\    \relax
[O~{\sc i}]                       & 6363.78 & $-$40.3$\pm$3.2       & $-$12.2$\pm$1.5   & 2.3$\pm$0.2     & 1.5$\pm$0.1 & -- & -- \\   \relax
H$\alpha$          & 6562.85 & $-$227.5$\pm$25.0  & $-$43.4$\pm$3.9   & 18.8$\pm$1.5   & 5.6$\pm$0.7 & $-$9.08$\pm$0.37 & $-$9.99$\pm$0.41  \\  \relax
[N~{\sc ii}]                      & 6583.45 & $-$10.3$\pm$2.1      & $-$4.2$\pm$0.3      & 0.9$\pm$0.1   & 0.7$\pm$0.1 & -- & -- \\   \relax
[S~{\sc ii}]                      & 6716.44 & $-$12.5$\pm$1.0      & $-$11.6$\pm$0.9    & 0.8$\pm$0.2       & 1.5$\pm$0.1 & -- & -- \\   \relax
[S~{\sc ii}]                      & 6730.82 & $-$18.3$\pm$2.9      & $-$17.7$\pm$1.4    & 1.3$\pm$0.1       & 2.5$\pm$0.2 & -- & -- \\ 
Ca~{\sc ii}                    & 8498.02 & $-$14.7$\pm$4.4      & $-$4.2$\pm$0.4       & 0.7$\pm$0.1    & 0.5$\pm$0.1 & $-$9.09$\pm$0.53 & $-$9.52$\pm$0.55 \\
Ca~{\sc ii}                    & 8542.09 & $-$13.9$\pm$1.9      & $-$3.9$\pm$0.3       & 0.9$\pm$0.1       & 0.5$\pm$0.04 & $-$9.18$\pm$0.59 & $-$9.63$\pm$0.62 \\
Ca~{\sc ii}                    & 8662.14 & $-$17.4$\pm$2.9      & $-$5.7$\pm$0.4       & 1.2$\pm$0.4       & 0.6$\pm$0.07 & $-$8.98$\pm$0.63 & $-$9.46$\pm$0.66 \\
\hline             
\end{tabular}
\end{table*}

\begin{figure}
     \includegraphics[width=92mm]{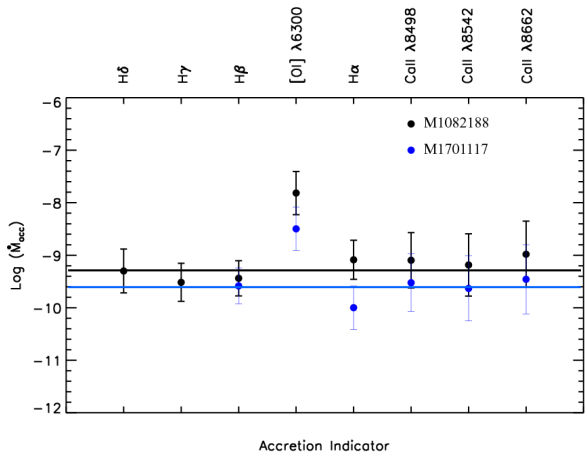}          
    \caption{The accretion rates $\log$ $\dot{M}_{acc}$ in $M_{\sun}$ yr$^{-1}$ derived from various line diagnostics (labelled) for M1082188 (black) and M1701117 (blue). Horizontal lines indicate the mean level calculated without including the [O~{\sc i}] line.  } 
    \label{accretion} 
\end{figure}

Table~\ref{fluxes} lists the observed pseudo-equivalent widths (pEWs) and line fluxes for the accretion and outflow emission lines. We estimate an uncertainty on the line luminosities and pEWs of $\sim$8--30\%, which arises from the noise level in the pseudo-continuum selected. We obtained estimates on the disc mass accretion rate, $\dot{M}_{acc}$, using multiple diagnostics of the Balmer and Ca~{\sc ii} IRT lines (Table~\ref{fluxes}), and the latest line luminosity relationships from Alcal\'{a} et~al. (2014). Figure~\ref{accretion} shows a plot for $\dot{M}_{acc}$ derived from these line indicators. The mean $\dot{M}_{acc}$ is measured to be (6.9$\pm$0.7)$\times$10$^{-10}$ $M_{\sun}$ yr$^{-1}$ and (2.3$\pm$0.3)$\times$10$^{-10}$ $M_{\sun}$ yr$^{-1}$ for M1082188 and M1701117, respectively. The uncertainty on $\dot{M}_{acc}$ has been calculated by propagating the error on the relationships of Alcal\'{a} et~al. (2014) and the error on the line flux measurements. The $\dot{M}_{acc}$ estimates are consistent with each other when the known accretion diagnostics are used. In addition, we also measured $\dot{M}_{acc}$ from the [O~{\sc i}] $\lambda$6300\,{\AA} line as an indirect accretion indicator, using the relation from Herczeg \& Hillenbrand (2008). The main assumption in using a jet line as an accretion indicator is that the [O~{\sc i}] emission line forms in an accretion powered outflow, due to which the [O~{\sc i}] line luminosities are correlated with the accretion rates. These $\dot{M}_{acc}$ estimates are about an order of a magnitude higher; for M1082188, the rate is (1.5$\pm$1.4)$\times$10$^{-8}$ $M_{\sun}$ yr$^{-1}$, and for M1701117, the rate is (6.8$\pm$6.5)$\times$10$^{-9}$ $M_{\sun}$ yr$^{-1}$. An argument for a higher [O~{\sc i}] derived $\dot{M}_{acc}$ can be that the [O~{\sc i}] line is forming above the envelope, and therefore should not be extincted to the same extent as a direct accretion tracer, which comes from inside the obscuring envelope. The [O~{\sc i}] line can thus be used as an indirect tracer of accretion, and is likely to provide a more reliable measure of $\dot{M}_{acc}$ rather than the known diagnostics for stars extincted by an envelope. 

For the outflow mass loss rate, a widely used method is to use the line luminosity of the [S~{\sc ii}] $\lambda$6731\,{\AA} line, as outlined in Hartigan et~al. (1995). The relations in Hartigan et~al. have been derived from shock models and are based on certain assumptions about the FEL critical density, the electron density, and the velocity of the outflow. We used a $V_{tan}$ of 50 km s$^{-1}$ for both sources, since this value is similar to the estimates obtained for some known Class II brown dwarf outflow sources (e.g., Whelan et~al. 2009). Due to the low spectral resolution of our data, we cannot obtain a robust estimate on $V_{tan}$ at present. The critical density n$_{cr}$ is set to 1.3$\times$10$^{4}$ cm$^{-3}$ from Hartigan et~al. (1995). Using these values, the mass outflow rate, $\dot{M}_{out}$, is calculated to be (1.0$\pm$0.5)$\times$10$^{-9}$ $M_{\sun}$ yr$^{-1}$ for M1082188, and (1.2$\pm$0.6)$\times$10$^{-9}$ $M_{\sun}$ yr$^{-1}$ for M1701117. The uncertainties were calculated in a manner similar to the uncertainty for the $\dot{M}_{acc}$ measurements.

We consider the measurements on accretion and outflow rates as crude estimates, mainly due to the uncertainty in the extinction correction that should be applied to the observed line fluxes. For the present case, we applied a correction for the interstellar reddening towards the targets, using the Galactic reddening and extinction calculator provided by IRSA\footnote{\url{http://irsa.ipac.caltech.edu/applications/DUST/}}. This service uses the Schlafly \& Finkbeiner (2011) Galactic reddening maps to determine the total Galactic line-of-sight absorption at a given position. We obtained an $A_{V}$ estimate of 0.8 mag and 1.2 mag for M1082188 and M1701117, respectively, over an area of 5-arcmin radius centred at the coordinates of the targets. These estimates are consistent with the reddening estimates for the $\sigma$~Orionis region ($A_{V}$ $\leq$ 1 mag; Lee 1968; B\'{e}jar et~al. 1999). There are no usable lines in the optical, such as the Pa $\delta$ and Br $\gamma$ lines observed in the near-infrared, which could provide a more robust estimate on the extinction in the emission line region for protostellar systems (e.g., Connelley \& Greene 2010). We plan to present in a subsequent study more robust estimates on the accretion and outflow activity rates using near-infrared spectroscopy.%, and using a spectro-astrometric analysis. 

%Finally, the extinction in the emission line region is expected to be higher than the interstellar reddening towards the sources, resulting in dereddened line fluxes higher than the observed values listed in Table~\ref{fluxes}. We can therefore treat the present accretion rates as lower limits. 

\section{Discussion}
\label{discuss}

We followed previous surveys on Class I protostars by e.g., Enoch et~al. (2009), Dunham et~al. (2008), Furlan et~al. (2008), and Evans et~al. (2009), where the bolometric luminosity of the source, as measured from integration of the observed SED over a wide range in wavelength from near-infrared to sub-millimetre/millimetre, has been used to reflect on the nature of the object as a very low-luminosity or a high-luminosity protostar. For our targets, the integrated bolometric luminosity, $L_{\rm bol}$, as measured from the observed SEDs covering $\sim$0.8--850\,$\mu$m, is 0.18$\pm$0.04 $L_{\sun}$ for M1701117, and 0.16$\pm$0.03 $L_{\sun}$ for M1082188. Considering an age of 3 Myr for $\sigma$~Orionis, these $L_{\rm bol}$ estimates correspond to a mass of $\sim$0.3$-$0.35\,$M_{\sun}$, based on the BT-Settl evolutionary models by Allard et~al. (2003), and is in the very low-mass range. The typical boundary between low-mass stars and very low-mass objects is considered to be at M $\sim$ 0.4\,$M_{\sun}$ (e.g., Chabrier \& Baraffe 1997; 2000). A further point in favour of the targets being very low-mass objects is that if these were T~Tauri stars, we would expect to see FELs with radial velocities of around 200 km\,s$^{-1}$. Jet radial velocities of $<$ 50 km\,s$^{-1}$ as we see for both sources are more typical of very low-mass stars and brown dwarfs (e.g., Whelan et~al. 2009). We can argue that since both targets are driving strong outflows, some of the envelope material might be dissipated further by the outflow entrainment, as has been suggested in recent studies on Class I outflow sources (e.g., Koyamatsu et~al. 2014; Hirano \& Liu 2014). Therefore, we do not discard the possibility that these objects are proto-brown dwarf candidates, and will probably form sub-stellar objects in the long run, reaching a final mass of the system close to or below the sub-stellar limit.

\begin{figure}  
     \includegraphics[width=80mm]{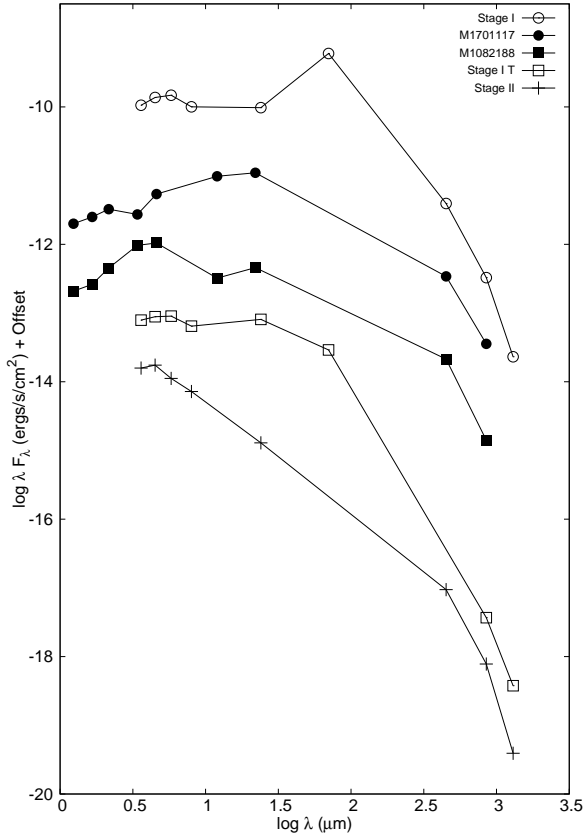} \\    
    \caption{The observed SED for M1701117 (filled circles) and M1082188 (filled squares) compared with the the SED for a Stage I (IRS 37), a Stage I-T (WL 17), and a Stage II (SR 21) source from van~Kempen et~al. (2009).  } 
    \label{SED} 
 \end{figure}

In Fig.~\ref{SED}, the observed SEDs for the targets are compared with those for a Stage I (IRS 37), a Stage I-Transition (Stage I-T; WL 17), and a Stage II (SR 21) source from van~Kempen et~al. (2009). These are tenuous envelope sources with low envelope masses of $<$ 0.1 $M_{\sun}$. The ``Stage'' classification scheme as first introduced by Whitney et~al. (2003) and later modified by van~Kempen et~al. (2009) is based on the physical characteristics of a YSO, such that Stage I objects have  0.1 $<$ $M_{disc}$/$M_{\rm env}$ $<$ 2, Stage II have $M_{\rm env}$=0, while Stage 3 are pre-main sequence stars with tenuous discs. The Stage I-T sources are in a transition from the embedded to pre-main sequence stage, and have tenuous envelopes compared to the early Stage I objects. The {\it best} model-fit estimates for our targets (Table~\ref{modelfit}) would classify both as Stage I objects. The SEDs of M1701117 and M1082188 in Fig.~\ref{SED} are flatter compared to the Stage II source, and more similar to the Stage I SED. A comparison of the SEDs for the two targets shows that M1082188 is a more evolved system than M1701117. What is also notable in Fig.~\ref{SED} is the ratio of the mid-infrared to the sub-millimetre fluxes, which is about four orders of magnitude for Stage I-T and Stage II SEDs, much larger than the targets or the Stage I source. As the envelope material dissipates, the cold dust mass in the system decreases, resulting in lower sub-millimetre fluxes and a large contrast between mid-infrared and sub-millimetre fluxes. The shape of the SED therefore becomes much steeper, compared to earlier stage SEDs. Therefore, both the observed and physical properties of the targets indicate an earlier evolutionary stage than typical Class II/Stage II sources. Also notable is the disc mass range for both sources, which is at least two orders of magnitude larger than the disc masses estimated from {\em Herschel}/PACS observations for Class II very low-mass stars and brown dwarfs at similar ages (Harvey et~al. 2012), and further indicates the earlier evolutionary stage for the targets.

Among low-mass Class I protostars in the Taurus and the Chamaeleon I and II star-forming regions ($\sim$1--3 Myr), the $\dot{M}_{acc}$ as measured from optical/near-infrared spectroscopy span a wide range between $\sim$1$\times$10$^{-7}$\,$M_{\sun}$ yr$^{-1}$ and 4$\times$10$^{-10}$ $M_{\sun}$ yr$^{-1}$, while the values for $\dot{M}_{out}$ derived from the [S~{\sc ii}] lines have about four orders of magnitude range between 2$\times$10$^{-10}$ and 3$\times$10$^{-6} M_{\sun}$ yr$^{-1}$ (e.g., White \& Hillenbrand 2004; Antoniucci et~al. 2011). For the few very-low luminosity Class I objects named IRAS 04158+2805, IRAS 04248+2612, and IRAS 04489+3042, with $L_{\rm bol}$ of $\sim$0.1--0.2\,$L_{\sun}$ identified in Taurus, the accretion and outflow rates range between $\sim$1$\times$10$^{-9}$\,$M_{\sun}$ yr$^{-1}$ and 4$\times$10$^{-10}$ $M_{\sun}$ yr$^{-1}$ (White \& Hillenbrand 2004). The $\dot{M}_{out}$ and the mean $\dot{M}_{acc}$ for our very low-mass Class I/Flat targets lie within this range. In comparison, the typical mass accretion rates for Class II very low-mass stars in young clusters at ages of $\sim$1--3 Myr are found to be small, with $\dot{M}_{acc}$ of the order of 10$^{-10}$ -- 10$^{-12} M_{\sun}$ yr$^{-1}$ (e.g., Muzerolle et~al. 2003; 2005), while the estimated range in $\dot{M}_{out}$ is $\sim$ (2--11)$\times$10$^{-10}$\,$M_{\sun}$ yr$^{-1}$ (e.g., Whelan et~al. 2009). Recent studies using high-resolution spectroscopy of strong accretors among Class II very low-mass stars in $\sigma$~Orionis have reported $\dot{M}_{acc}$ between $\sim$4$\times$10$^{-10}$\,$M_{\sun}$ yr$^{-1}$ and 4$\times$10$^{-11}$ $M_{\sun}$ yr$^{-1}$, with the exception of the Class II object Mayrit 1040182 (V604\,Ori), which is an intense accretor with $\dot{M}_{out}$ of $\sim$1$\times$10$^{-9}$\,$M_{\sun}$ yr$^{-1}$ (Rigliaco et~al. 2012). Overall, these comparisons suggest that the activity rates are not significantly different between Class I and Class II stages for very low-mass stars, and are also within the wide range observed for the Class I low-mass stars. This has also been noted by e.g., White \& Hillenbrand (2004), who found the median $\dot{M}_{acc}$ and $\dot{M}_{out}$ for Class I and Class II low-mass stars to be indistinguishable. The observed trends can be explored further once the outflow and accretion properties for a greater number of very low-mass stars during the early evolutionary stages have been studied, and the rates can be calculated with less degeneracy.

\section{Summary}

We conducted a multi-wavelength study of two very low-luminosity Class I/Flat sources, Mayrit 1701117 and Mayrit 1082188, in the $\sigma$ Orionis cluster. Both objects exhibit prominent signatures of accretion and outflow activity, with the mean accretion rate of 6.4$\times$10$^{-10}$ $M_{\sun}$ yr$^{-1}$ for M1701117, and 2.5$\times$10$^{-10}$ $M_{\sun}$ yr$^{-1}$ for M1082188. The outflow mass loss rates for the two systems are similar and estimated to be $\sim$1$\times$10$^{-9}$\,$M_{\sun}$ yr$^{-1}$. The activity rates lie within the range observed for low-mass Class I protostars. The bolometric luminosity of the targets as measured from the observed spectral energy distribution over $\sim$0.8--850\,$\mu$m is 0.18$\pm$0.04 $L_{\sun}$ for Mayrit 1701117, and 0.16$\pm$0.03 $L_{\sun}$ for Mayrit 1082188. The total dust+gas mass from the envelope+disc components, derived from the 850\,$\mu$m flux, is estimated to be $\sim$36\,$M_{\rm Jup}$ and $\sim$22\,$M_{\rm Jup}$ for Mayrit 1701117 and Mayrit 1082188, respectively. There is the possibility that some of the envelope material might be dissipated by the strong outflows driven by these sources, resulting in a final mass of the system close to or below the sub-stellar limit. We are conducting further follow-up observations at various wavelengths to perform a more detailed charaterization study, which will help shed light on the true nature of these sources.

\section*{Acknowledgments}

We thank B. Whitney, L. Hillenbrand, M. Connelley, M. R. Zapatero Osorio, M. Bate, T. Robitaille, F. Allard, and J. Hern\'{a}ndez for insightful discussions, and A. Chrysostomou for help with the SCUBA-2 data reduction. E. T. Whelan acknowledges financial support from the Deutsche Forschungsgemeinschaft through the Research Grant Wh 172/1-1. N. Lodieu is funded by the Ram\'on y Cajal fellowship number 08-303-01-02. This project was partly funded by the national program AYA2010-19136 funded by the Spanish ministry of Science and Innovation. The James Clerk Maxwell Telescope is operated by the Joint Astronomy Centre on behalf of the Science and Technology Facilities Council of the United Kingdom, the Netherlands Organisation for Scientific Research, and the National Research Council of Canada. Additional funds for the construction of SCUBA-2 were provided by the Canada Foundation for Innovation. Based on observations collected at the Centro Astron\'{o}mico Hispano Alem\'{a}n (CAHA) at Calar Alto, operated jointly by the Max-Planck Institut f\"{u}r Astronomie and the Instituto de Astrof'sica de Andaluc'a (CSIC).

\label{lastpage}

\end{document}